# Sparse learning enabled by constraints on connectivity and function

Mirza M. Junaid Baig[1,2] and Armen Stepanyants[1]

[1]Department of Physics, Center for Theoretical Biological Physics, and [2]Department of Bioengineering, Northeastern University, Boston, MA

**Corresponding author:** Armen Stepanyants; email: a.stepanyants@northeastern.edu

## ABSTRACT

Sparse connectivity is a hallmark of the brain and a desired property of artificial neural networks. It promotes energy efficiency, simplifies training, and enhances the robustness of network function. Thus, a detailed understanding of how to achieve sparsity without jeopardizing network performance is beneficial for neuroscience, deep learning, and neuromorphic computing applications. We used an exactly solvable model of associative learning to evaluate the effects of various sparsity-inducing constraints on connectivity and function. We determine the optimal level of sparsity achieved by the $\ell_0$ norm constraint and find that nearly the same efficiency can be obtained by eliminating weak connections. We show that this method of achieving sparsity can be implemented online, making it compatible with neuroscience and machine learning applications.

## INTRODUCTION

Sparsity is a central theme across various disciplines. In neuroscience, it is a characteristic feature of brain networks [1], offering numerous advantages such as simplifying circuit development, reducing brain volume and wiring length, lowering metabolic cost, and streamlining the learning process. In the field of compressed sensing, sparsity enables the recovery of signals from a limited number of measurements [2]. In neuromorphic computing, sparsity is desired to reduce latency and power consumption and improve scalability and robustness [2]. In machine learning, sparsity is known to improve the generalizability and interpretability of the model and reduce computational cost and training time [3]. Across all these fields, sparsity simplifies the construction and maintenance of the systems, albeit at the expense of functionality, as fewer connections can simply do less. Thus, understanding how to achieve sparsity efficiently—without significantly compromising the system's function—is of paramount importance.

To study the tradeoff between sparsity and functionality in detail, we considered a model of associative memory storage by a single perceptron [4,5]. This model is inspired by the brain and serves as a building block in machine learning networks (Figure 1A). It offers the advantage of being analytically tractable while capturing key properties of larger networks, regarding learning capacity and connectivity. In its simplest form, the model was first solved by Cover [6], who employed a clever geometric argument to demonstrate that a perceptron with $N$ inputs can successfully learn the task of binary classification of up to $2N$ random input patterns. Subsequently, advances in statistical physics, such as replica theory and cavity method [7-9], have led to solutions of more general perceptron models [10-16], albeit in the $N \to \infty$ limit. In parallel with these efforts, the perceptron learning rule has been developed to provide numerical solutions for finite learning problems, ensuring convergence after a fixed number of steps in feasible cases [17-19]. This rule can be used in online settings, where learning examples are presented sequentially, making it highly relevant for both neuroscience and machine learning applications. However, findings from replica theory and applications of the perceptron learning rule have revealed that the relatively high learning capacity of the simple perceptron is achieved through dense connectivity.

The central message of this study is that sparsity in learning arises from constraints on connectivity and function. For example, one may create sparsity by simply cutting some connections before learning, a method known as quenched $\ell_0$ norm constraint ($q\ell_0$). However, this naïve approach is inefficient because, without prior knowledge of the learning examples, it is impossible to determine which connections to cut. On the other hand, the annealed $\ell_0$ norm constraint yields the optimal way of inducing sparsity. Unfortunately, it results in an NP-hard problem [20] that cannot be efficiently solved in an online learning environment. Nonetheless, as we demonstrate below, an analytical solution exists in the $N \to \infty$ limit, providing a valuable benchmark for evaluating other methods.

In this study, we explore several additional methods of inducing sparsity during learning by imposing constraints, focusing on those inspired by the brain. We show how the strengths of the constraints mediate tradeoffs between sparsity and learning capacity. Notably, we introduce *gap* constraints that, by disallowing small-weight connections, yield a near-optimal tradeoff between sparsity and learning capacity, similar to that achieved by the $\ell_0$ norm. We develop a perceptron-type learning rule that enables online associative learning under *gap* constraints and show that, in addition to promoting sparsity, these constraints enhance both accuracy and generalizability.

**Associative memory storage with sparsity-inducing constraints (SICs)**

To study the effects of SICs on network connectivity and function we considered a model of associative memory storage by a single perceptron (Figure 1A). In its most basic form, the model works as follows. A perceptron of $N$ inputs (enumerated with index $j$) must store $m$ binary (0, 1) input-output associations $X^\mu \to y^\mu$, $\mu = 1,...,m$, by adjusting its input connection strengths, $J_j$. Components of vectors $X^\mu$ ($X_j^\mu$) and scalars $y^\mu$ are randomly and independently drawn from Bernoulli distributions in which the probabilities of having a 1 are denoted with $f_j$ and $f_{out}$ respectively. An association $\mu$ with $y^\mu = 0$ (or 1) is said to be successfully stored if the postsynaptic input to the perceptron, $\sum_{j=1}^{N} J_j X_j^\mu$, is < 0 (or > 0). These inequalities can be combined with the help of the Heaviside step function.

$$\theta\left(\sum_{j=1}^{N} J_j X_j^{\mu}\right) = y^{\mu}, \quad \mu = 1,\ldots,m \tag{1}$$

The properties of this basic model have been extensively studied analytically and numerically. Figure 1B shows that the probability of storing a set of $m$ associations in a perceptron is a decreasing function of the memory load, $m/N$. With increasing load, this probability undergoes a transition from 1 to 0 as the learning problems become increasingly unfeasible. The memory load corresponding to the success probability of 0.5 is referred to as the perceptron's associative memory storage capacity, $\alpha$. With increasing $N$, the transition from successful learning to inability to learn the entire set of associations becomes sharper, and $\alpha$ tends to its $N \to \infty$ limit referred to as the critical capacity, $\alpha_c$.

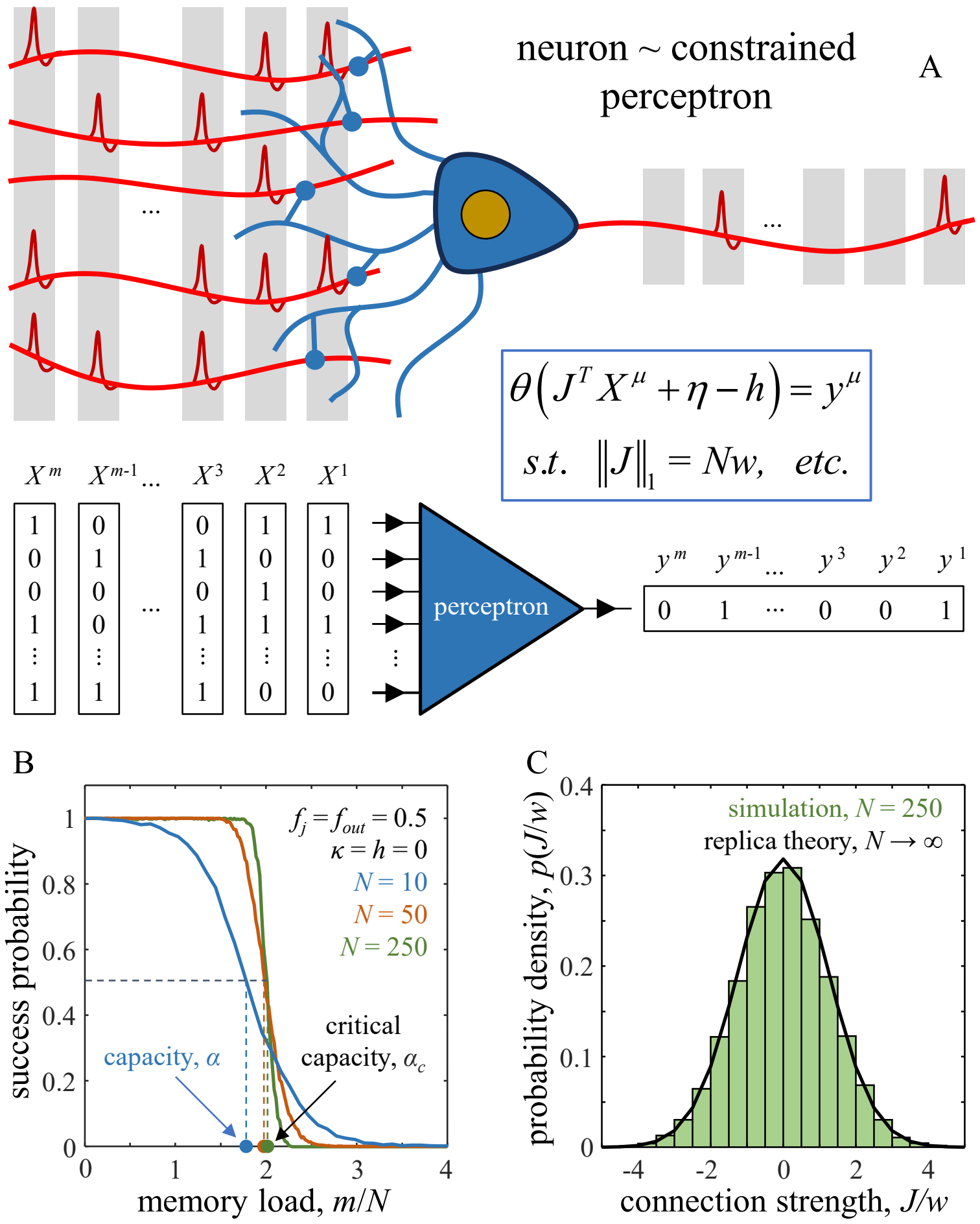

**Figure 1:** Constrained perceptron as a neuron-based model of associative learning. **A, top**. Schematic of a cortical neuron receiving presynaptic inputs from multiple axons (red) carrying action potentials. The neuron generates output action potentials when the postsynaptic input exceeds its threshold of firing, $h$. **A, bottom.** The neuron is paralleled by a binary perceptron receiving inputs $X^\mu$ and producing outputs $y^\mu$ at discrete times. The perceptron can learn a set of such associations by adjusting its connection strengths, $J$, in the presence of various biologically inspired constraints. **B**. The probability of successful learning of a set of associations is a decreasing function of the number of associations, $m$. The transition from perfect learning to inability to learn the entire set becomes progressively sharper with an increasing number of perceptron inputs, $N$. In this process, the capacity of the perceptron for associative memory storage, defined as the memory load corresponding to 50% success probability, approaches its critical value that can be determined with the replica theory in the $N \to \infty$ limit. **C**. At capacity, the distribution of connection strengths of a perceptron receiving $N$ = 250 inputs (green bars) is indistinguishable from that obtained with replica theory (black line). Results were obtained for an $\ell_1$ norm constrained perceptron with model parameters shown in (B, C).

Because the solution space of Eq. (1) is open (if $\{J_j\}$ is a solution, so is $\{cJ_j\}$, $\forall\ c > 0$), we had to regularize this basic model to study the perceptron's connection strengths. We did this by fixing the total magnitude of connection strengths, $\sum_{j=1}^{N}\left|J_j\right| = Nw$, where $w$ represents the average connection weight. This regularization is inspired by the brain where the strength of a synaptic connection is known to be correlated with such limiting resources as the number of vesicles in the presynaptic bouton and volume of the postsynaptic dendritic spine [21]. The $\ell_1$ norm regularization is ubiquitously used to produce sparse solutions in regression, compressed sensing, and deep learning applications. Here, however, this constraint by itself does not induce sparse connectivity during learning. Our analytical and numerical results show that at capacity the probability density of the perceptron's connection strengths is Gaussian (Figure 1C and Appendix), and the associations are stored with dense connections.

We investigated the effects of five brain-inspired constraints, added to the basic model to induce sparsity during learning. (1) Since neurons in the cerebral cortex are thought to operate with fixed firing thresholds [13], we equipped the basic model of Eq. (1) with a fixed threshold, $h \geq 0$. (2) It is reasonable to assume that memory retrieval in the brain must tolerate a certain level of postsynaptic noise [22]. Therefore, we considered the effects of a constraint that ensures perfect recall of stored associations in the presence of bounded noise $\eta$, $\left|\eta\right| \leq \kappa$, where $\kappa$ is referred to as the robustness parameter. (3) The strengths of synaptic connections in the cerebral cortex can change during learning, but an excitatory connection ($g_j$ = +1) generally does not become inhibitory ($g_j$ = –1) and vice versa (Dale's principle [23]). Therefore, we also examined memory storage with connections of fixed signs, which we refer to as *sign* constraints. (4) The fourth constraint keeps the number of non-zero-weight connections fixed during memory storage and is referred to as the $\ell_0$ norm constraint. Mathematically, this constraint can be expressed as $\lim_{\varepsilon \to 0}\sum_{j=1}^{N}\left|J_j\right|^{\varepsilon} = Np$, where $p$ denotes the fraction of non-zero-weight connections. Though there is no direct evidence that neurons in the brain maintain the total number of synapses during learning, we examined this constraint because it provides the optimal way of inducing sparsity and serves as a benchmark for other models. (5) Owing to its discrete molecular machinery, a functional synapse in the brain has a minimal weight, known as the quantal amplitude [24,25]. Therefore, we examined learning with connection strengths that must either be zero (no connection or silent synapse) or have minimal magnitudes (functional synapse), $|J_j| \geq \Delta_j \geq 0$. We refer to these restrictions as *gap* constraints as they create gaps in the connection strength distributions. A sparse associative learning model may include some or all these constraints:

$$\theta\left(\sum_{j=1}^{N} J_j X_j^{\mu} - h + \eta\right) = y^{\mu}, \quad \mu = 1,\ldots,m$$

subject to:

$$\begin{cases} \ell_1: & \sum_{j=1}^{N} |J_j| = Nw \\ h: & h - \text{constant} \\ \kappa: & |\eta| \leq \kappa \\ sign: & J_j g_j \geq 0, \quad j = 1,\ldots,N \\ \ell_0: & \lim_{\varepsilon \to 0} \sum_{j=1}^{N} |J_j|^{\varepsilon} = Np \\ gap: & |J_j| \geq \Delta_j \geq 0 \quad if \quad J_j \neq 0 \end{cases} \tag{2}$$

To provide intuition for why the selected constraints lead to sparsity during learning, we note that the problem of memory storage to capacity, i.e., maximization of memory load for a fixed *w*, is equivalent to the problem of minimization of *w* for a given load. The solution region to the former problem is not easy to visualize as it involves an increasing number of associations, each represented by a hyperplane that selects half of the connection strength space. The solution region of the latter problem can be visualized more easily as it involves a continuously changing variable *w*. Figure 2 illustrates the solution region for a perceptron of $N = 2$ inputs (axes) learning $m = 2$ associations (blue half-planes) in the presence of *h*, *κ*, $\ell_0$, or *gap* constraints. In these cases, minimization of *w* selects a unique solution that is sparse.

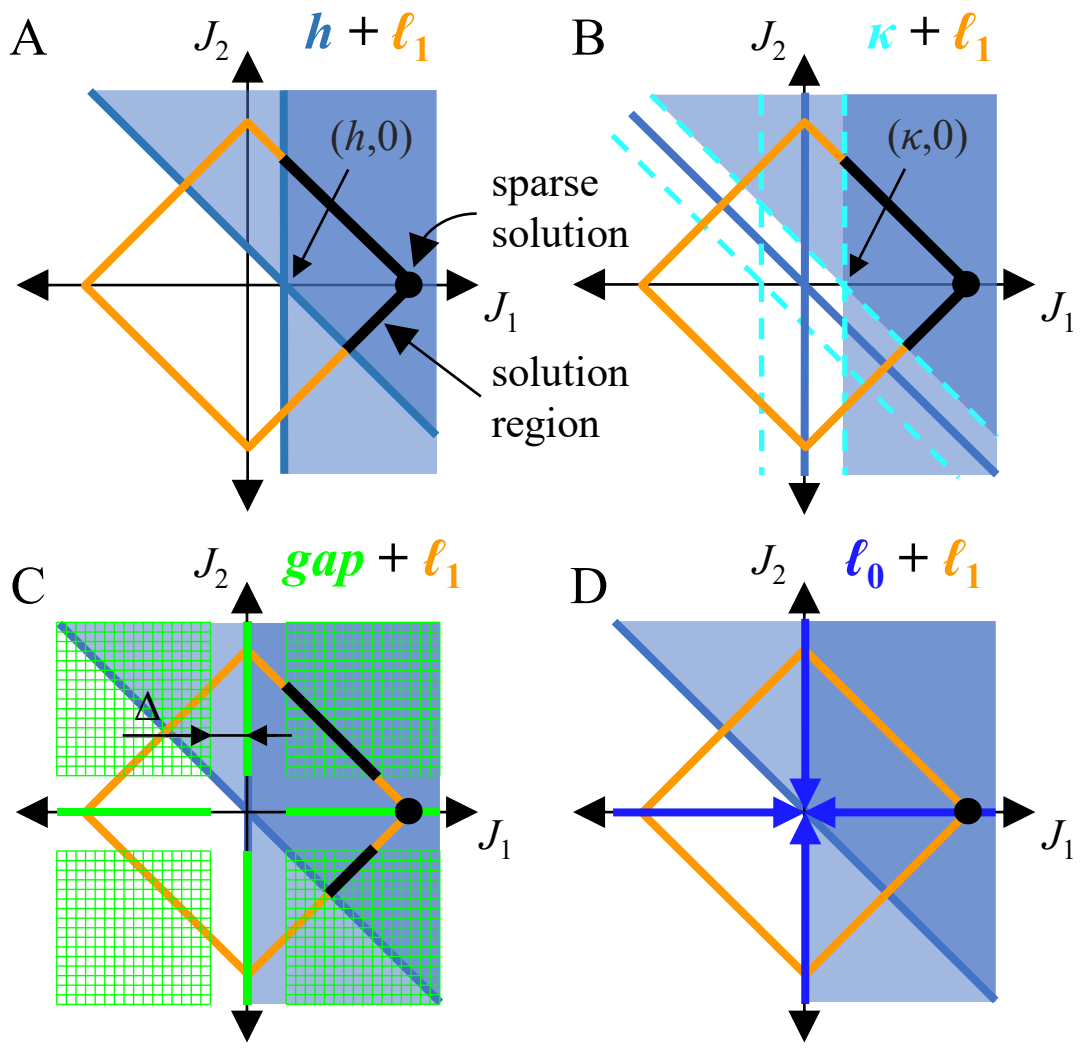

**Figure 2:** Constraints can induce sparse connectivity. **A.** A binary perceptron with $N$ = 2 inputs and a threshold $h$ > 0 is loaded with $m$ = 2 associations (blue half-planes) in the presence of an $\ell_1$ norm constraint (orange square). The solution region of this problem is indicated with a black line. Minimization of $\ell_1$ norm (or maximization of $h$), leads to a unique solution that is sparse (black dot). Similar arguments hold in the presence of robustness constraint, $\kappa$ (**B**), and *gap* constraints, Δ (**C**). **D.** In the presence of an $\ell_0$ norm constraint (blue arrows, $p$ = 0.5 in example) the solution is sparse by design.

## Analytical solution of the sparse learning model in the large *N* limit

Because in typical brain and artificial networks, neurons receive large numbers of inputs, $N \sim 10^3 - 10^6$, we first determined the solution of the sparse learning model of Eqs. (2) in the $N \to \infty$ limit which was done with the replica method. Our previous results and Figures 1B and 1C show that at capacity the replica solutions approximate well the properties of finite learning models for $N$ as low as ~100. Therefore, we extended the previously developed replica methods [10-13,16] and derived a general analytical solution of the sparse learning model that encompasses all combinations of constraints of Eqs. (2) (see Appendix).

Figure 3 illustrates how the critical capacity, $\alpha_c$, sparsity, $S$ defined as the fraction of zero-weight connections, and probability density of connection strengths, $p(J/w)$, depend on the five SIC strengths. The $\ell_1$ norm constraint was included in all cases to ensure that the solution remains finite, and all relevant quantities were normalized with $w$. For brevity, we only show the results for the homogeneous models with $f_j = f_{out} = 0.5$ and $\Delta_j = \Delta$. At its maximum, $\alpha_c = 2$ in agreement with the result of Cover [6]. This maximal capacity can only be attained in the absence of SICs by a fully connected perceptron, $S = 0$. The constraints increase $S$ at the expense of $\alpha_c$, establishing tradeoffs between these structural and functional properties of the perceptron.

At critical capacity, the probability density of connection strengths of a constrained perceptron, $p(J/w)$, is generally composed of two truncated Gaussians (one for connections with $J > 0$ and the other for $J < 0$) and a finite fraction of $J = 0$ connections (Figure 3, bottom row). In the $\ell_0 + \ell_1$ case, the connection strength distribution contains a gap on each side of $J = 0$, indicating the absence of weak non-zero-weight connections. This gap is induced by the $\ell_0$ norm, which eliminates weak connections to promote sparsity at minimal detriment to the perceptron's memory storage capacity. Since the $\ell_0$ constraint attains the optimal sparsity for a given $\alpha_c$ while inducing a gap in $p(J/w)$, we expected to achieve a similar result with an engineered gap. The solution of the model in the *gap* + $\ell_1$ case (Figure 3E) shows that this is indeed the case and that local *gap* constraints can promote sparsity nearly as well as the global $\ell_0$ constraint. In the *gap* case, however, $p(J/w)$ contains additional finite fractions of connections at the outer edges of the gaps, $S_\pm$, which is indicative of the slightly lower efficiency of this constraint in comparison with $\ell_0$.

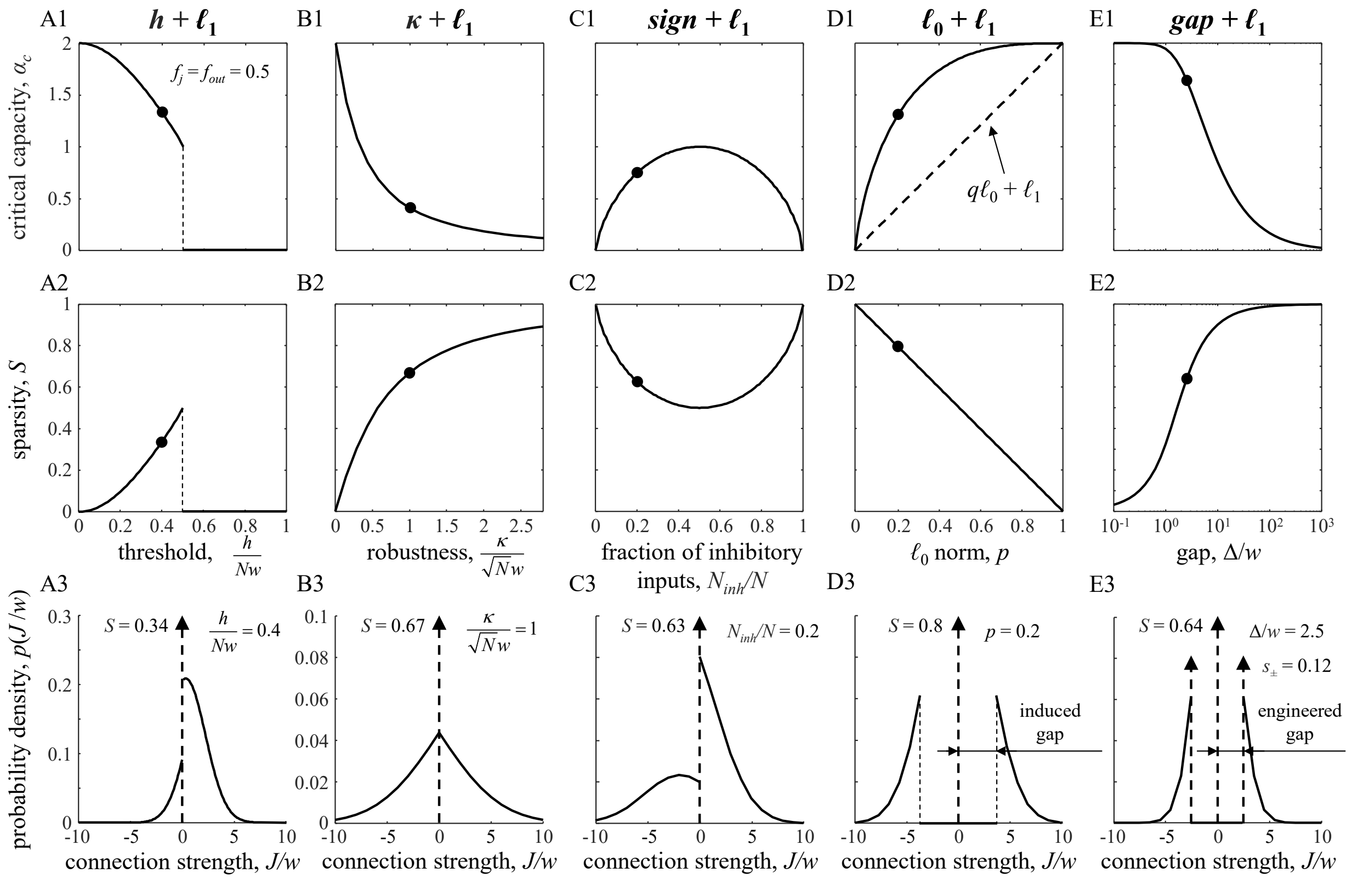

**Figure 3**: Effects of constraints on critical capacity and connectivity. Critical capacity (**A1**) and sparsity (**A2**) obtained with replica theory in the $h + \ell_1$ case are plotted as functions of threshold. Model parameters are displayed in (A1). **A3.** Probability density of connection strengths for $h/Nw = 0.4$ (black dots in A1,2) contains $S = 0.34$ fraction of zero-weight connections (dashed arrow at 0 representing a Dirac delta function). Same for the $\kappa + \ell_1$ (**B1-3**) and *sign* + $\ell_1$ (**C1-3**) cases. **D1,2.** Critical capacity and sparsity in the $\ell_0 + \ell_1$ case as functions of $p$. Dashed line in (D1) corresponds to the case of quenched $\ell_0$ norm constraint ($q\ell_0$). **D3.** Probability density of connection strengths contains gaps on both sides of zero. **E1,2.** Critical capacity and sparsity in the *gap* + $\ell_1$ case as functions of the gap size. **E3.** Probability density of connection strengths in the case of $\Delta/w = 2.5$ contains $S = 0.64$ fraction of zero-weight connections and $s_\pm = 0.12$ fractions of connections at the outer edges of the gaps.

## The tradeoff between capacity and sparsity is mediated by constraints

Figure 4A explicitly shows the tradeoff between critical capacity and sparsity in the five constraint cases. The $\ell_0 + \ell_1$ constraint (blue line) provides the optimal tradeoff, i.e., it is the closest curve to the unattainable $S = 1$, $\alpha_c = 2$ point (top right corner). Remarkably, this constraint makes it possible to achieve nearly maximal capacity ($\alpha_c = 1.86$) with only 50% of the input connections ($S = 0.5$). The *gap* + $\ell_1$ case closely follows the optimal tradeoff line, achieving $\alpha_c = 1.82$ with 50% of connections. This is much better than the trivial way of achieving sparsity by simply pruning a set fraction of connections before learning ($q\ell_0$, thick gray line). The latter is equivalent to the $h + \ell_1$ and *sign* + $\ell_1$ cases, which, however, have more limited ranges. The $\kappa + \ell_1$ case yields the least effective tradeoff between capacity and sparsity. However, by directly enforcing the robustness of stored memories, this constraint provides an added functional benefit.

Notably, the two best tradeoff cases, $\ell_0 + \ell_1$ and *gap* + $\ell_1$, function with no weak non-zero-weight connections. This is not unexpected, as eliminating a weak connection has a relatively small effect on the perceptron's function while increasing sparsity by the same fixed amount, $1/N$, removal of any other connection would produce. In the former case, the gaps on both sides of zero in the connection strength distribution (Figure 3D3) are induced by the $\ell_0$ constraint, while in the latter these gaps are explicitly engineered into the constraint (Figure 3E3). Figure 4B shows that in both cases sparsity is correlated with the gap size, $\Delta$. With $\ell_0 + \ell_1$, $S = 0.5$ is associated with a gap of $\Delta = 1.06w$, while in the *gap* + $\ell_1$ case, a larger gap, $1.66w$, is required to achieve the same level of sparsity consistent with the lower efficiency of this method.

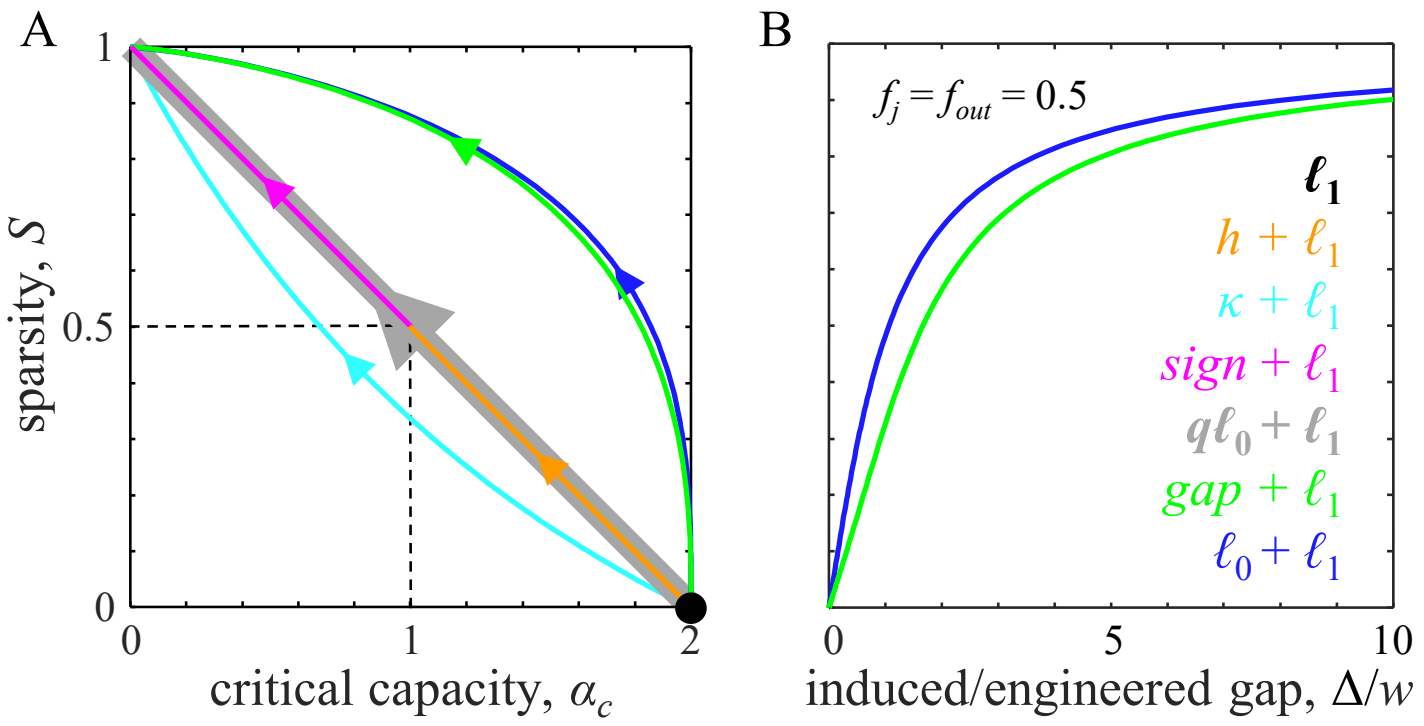

**Figure 4:** Sparsity in constrained perceptron models. **A.** Tradeoff between sparsity ($S$) and critical capacity ($\alpha_c$) in constrained perceptron models (key in B). Arrows indicate the directions of increasing constraint strengths. Gray line, corresponding to the $q\ell_0 + \ell_1$ case, is thickened for better visibility. **B.** In the $\ell_0 + \ell_1$ case, sparsity induces a gap in connection strength distribution (blue), while in the *gap* + $\ell_1$ case (green) gap is engineered to induce sparsity. Results in (A, B) were obtained with the replica theory for model parameters displayed in (B).

## Online learning with SICs

Next, we set out to test if the intuition gained from the above analytical solution extended to more practical problems such as learning to discriminate finite and correlated input patterns online. Such problems are encountered in machine learning applications and are arguably faced by neurons in the brain. They are generally unfeasible and only require "good enough" or approximate solutions that produce correct results in most cases. In online learning, examples are presented one at a time, and learning proceeds stochastically to an approximate solution for a certain number of steps or until the desired accuracy is achieved. To that end, we modified the perceptron learning rule [18] so that, in addition to learning the associations, every step of the algorithm would attempt to enforce the SICs of Eqs. (2) (see Appendix for details). Due to the NP-hard nature of the $\ell_0$ norm constraint, we did not attempt to extend the sparse learning rule on the $\ell_0 + \ell_1$ case but considered $q\ell_0 + \ell_1$ in its place.

We evaluated the performance of the sparse learning rule on the task of classification of handwritten digits of the MNIST dataset [26]. A constrained perceptron model was trained to classify each digit. Each model was trained on 60,000 and tested on 10,000 examples in which all images of that digit, $X^\mu$, were associated with $y^\mu = 1$ and the rest with $y^\mu = 0$. Because the binary classification problems solved by these models are unfeasible and highly unbalanced (Positive/Negative examples ~ 1/9 ratio), Balanced Accuracy [27] was used instead of capacity to evaluate their performance on training and testing examples, and the results of all models were averaged.

Figure 5 shows that the results obtained with the sparse learning rule are in qualitative agreement with the analytical results of Figures 3 and 4. The differences could be attributed to the fact that the MNIST images are spatially correlated, while replica results were derived for random input patterns. Figure 5 illustrates that the highest balanced accuracy during training is achieved in the absence of SICs with dense connections, $S = 0$. With increasing strengths of the constraints, the accuracy declines monotonically while $S$ rises, leading to a tradeoff between these quantities, analogous to the analytical results of Figure 3. All SICs improve the generalizability of the models evidenced by the convergence of the training and testing accuracy curves with increasing strengths of the constraints. As expected, robustness to noise can increase testing accuracy as illustrated by the initial rise of the green curve in Figure 5B1. Interestingly, the same trend was observed in the *gap* + $\ell_1$ case, demonstrating that the removal of weak connections, in addition to increasing generalizability, can improve the model's robustness to noise and small variations in examples.

Figure 5 (bottom row) shows that the tradeoff lines of balanced accuracy and sparsity are arranged in an order similar to the analytical results of Figure 4A. In the absence of the $\ell_0 + \ell_1$ case, the best tradeoff is achieved by the *gap* constraints, Figure 5E3. A balanced test accuracy of 0.92 is attained by an unconstrained system with dense connections, $S = 0$. Remarkably, the same test accuracy could be achieved in the *gap* case with only 15% of connections, and the maximum test accuracy of 0.94 required 40% of connections.

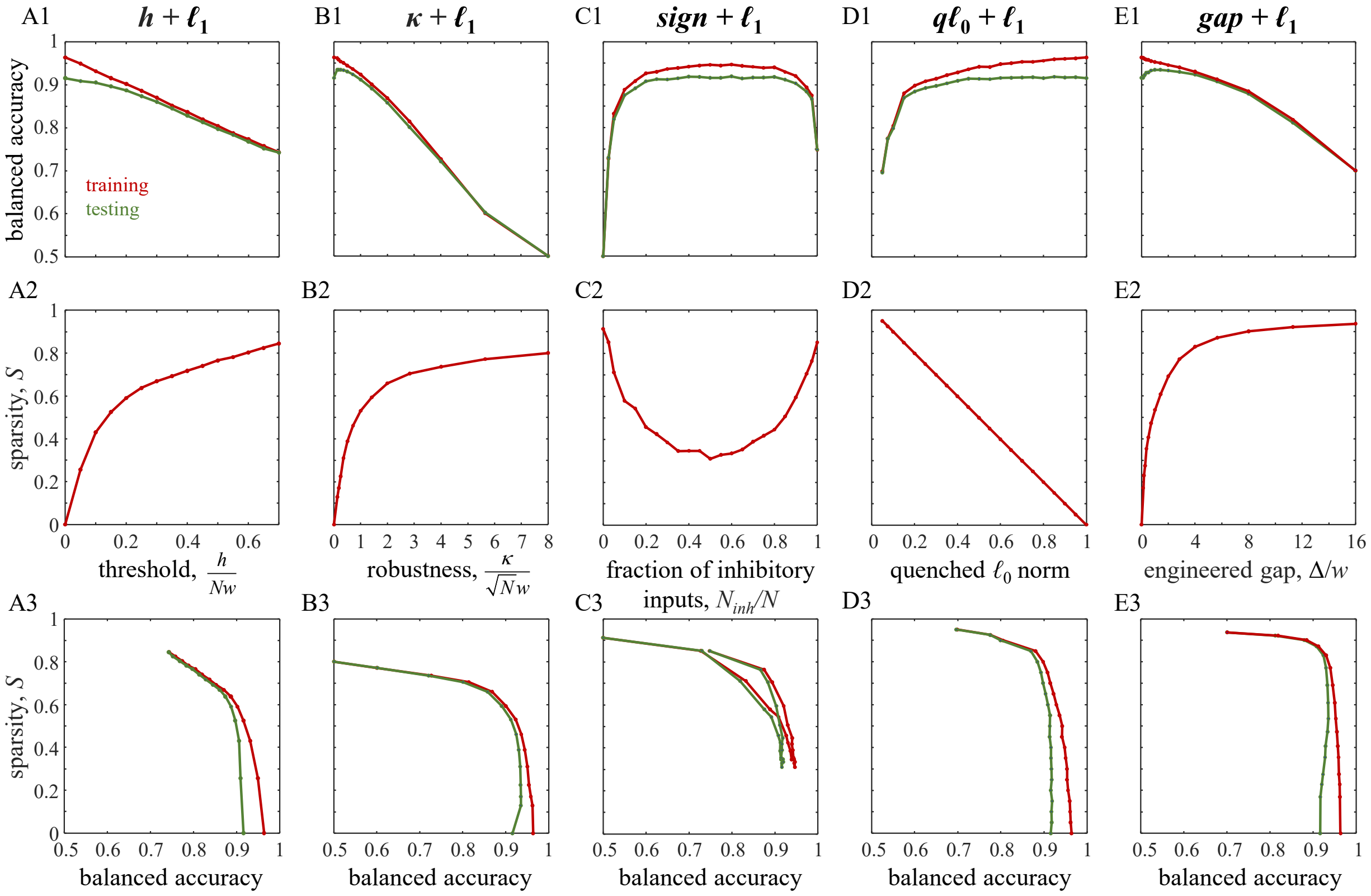

**Figure 5**: Sparse learning can be done online with a perceptron-type learning rule. Balanced accuracy (**A1**) and sparsity (**A2**) after training on the MNIST dataset of handwritten digits in the $h + \ell_1$ constrained case are plotted as functions of robustness. The model was trained for $10^6$ learning steps with the sparse learning rule. The training (red) and testing (green) results were averaged over all digits. **A3.** Same in the $\kappa + \ell_1$ (**B1-3**), *sign* + $\ell_1$ (**C1-3**), $q\ell_0 + \ell_1$ (**D1-3**), and *gap* + $\ell_1$ (**E1-3**) constrained cases. The tradeoff between sparsity and balanced accuracy parallels the theoretical results of Figure 4A.

## DISCUSSION

We derived analytical solutions for a class of learning models that incorporate brain-inspired constraints on connectivity and function. Our results revealed how these constraints mediate the tradeoffs between sparsity and learning capacity. We determined the optimal tradeoff level induced by the $\ell_0$ norm constraint and demonstrated that $\ell_0$ leads to the absence of weak connections. A similar feature is observed in the brain, where, due to its discrete components, a functional synapse must have a minimum non-zero weight. Drawing inspiration from these insights, we introduced *gap* constraints and showed that they are nearly as effective as $\ell_0$ in inducing sparsity with minimal detriment to learning capacity.

The $\ell_0$ norm constraint, due to its global and non-convex nature, is known to result in NP-hard problems, for which efficient numerical solutions are not available. While the complexity of the gap-constrained problem is not known, it appears to be more amenable to approximate numerical algorithms that can be implemented online, presumably because constraints are applied to individual connections. We demonstrated this by developing a perceptron-type learning rule for finding approximate solutions to constrained learning problems and applying it to the classification of handwritten digits. Figure 5E3 shows that *gap* constraints enhance sparsity, improve testing accuracy, and reduce overfitting. It remains to be seen if *gap* constraints can be effectively implemented in deep learning applications.

Although a direct quantitative comparison of learning in the brain and the basic learning model presented here is unwarranted, we provide some numerical values to stimulate further research. It is well established that the brain operates with sparse connectivity, with $S \approx 0.8$-$0.9$ in the neocortex [16]. The quantal amplitude can be estimated from paired recordings combined with the synaptic transmission model of del Castillo & Katz [24], yielding $\Delta/w \approx 0.2$-$0.4$ [25]. In contrast, the maximum testing accuracy in Figure 5E1 corresponds to $S = 0.55$ and $\Delta/w = 1.1$, and to achieve the level of sparsity observed in the brain (e.g., $S = 0.85$), a much larger gap is required, $\Delta/w = 4.8$. This discrepancy is likely to be reduced by combining several SICs of Eqs. (2), which would provide a somewhat more appropriate model for brain neurons.

**ACKNOWLEDGMENTS**

This work was supported by AFOSR grant FA9550-15-1-0398, NSF grant IIS-1526642, and NIH grant R56 NS128413.

**APPENDIX**

This document summarizes the main steps of analytical and numerical solutions of the model described in the main text. The analytical solution, obtained in the large $N$ limit, is based on the replica method from statistical physics [7-10] and follows previously described calculation steps [12,13]. It extends the methodology developed by Bouten et al., [11] for the $\ell_0$ norm constraint on the cases of *sign* and *gap* constraints. A single general solution was derived that encompasses fixed threshold, robustness, $\ell_0$, $\ell_1$, *sign*, and *gap* constraints and combinations thereof. To validate this analytical result, the model was cast into a mixed-integer linear optimization problem and solved in MATLAB. A perceptron-type learning rule was used in combination with the MNIST dataset of handwritten digits [26] to show that consistent with the analytical results constraints can also lead to sparsity in online applications.

**Standardization of the model for replica and numerical solutions**

The following perceptron-based associative memory storage model with a fixed threshold, robustness, and four types of local and global constraints on connection strengths is described in the main text.

$$\theta\left(\sum_{j=1}^{N} J_j X_j^\mu - h + \eta\right) = y^\mu, \quad \mu = 1,\ldots,m$$

$$\text{Prob}\left(X_j^\mu\right) = \begin{cases} 1 - f_j, & X_j^\mu = 0 \\ f_j, & X_j^\mu = 1 \end{cases}; \quad \text{Prob}\left(y^\mu\right) = \begin{cases} 1 - f_{out}, & y^\mu = 0 \\ f_{out}, & y^\mu = 1 \end{cases}$$

subject to:

$$\begin{cases} \ell_1: & \sum_{j=1}^{N} \left|J_j\right| = Nw \\ h: & h - \text{constant} \\ \kappa: & |\eta| \le \kappa \\ sign: & J_j g_j \ge 0, \quad j = 1,\ldots,N \\ \ell_0: & \lim_{\varepsilon \to 0} \sum_{j=1}^{N} \left|J_j\right|^\varepsilon = Np \\ gap: & \left|J_j\right| \ge \Delta_j \ge 0 \quad if \quad J_j \ne 0 \end{cases} \tag{A1}$$

As a first step, the association equations were rewritten as inequalities that ensure accurate memory retrieval in the presence of any noise $\eta$ bounded by the robustness parameter $\kappa$.

$$\left(2y^\mu - 1\right)\left(\sum_{j=1}^{N} J_j X_j^\mu - h\right) > \kappa, \quad \mu = 1,\ldots,m$$

$$\text{Prob}\left(X_j^\mu\right) = \begin{cases} 1 - f_j, & X_j^\mu = 0 \\ f_j, & X_j^\mu = 1 \end{cases}; \quad \text{Prob}\left(y^\mu\right) = \begin{cases} 1 - f_{out}, & y^\mu = 0 \\ f_{out}, & y^\mu = 1 \end{cases}$$

subject to:

$$\begin{cases} \ell_1: & \sum_{j=1}^{N} \left|J_j\right| = Nw \\ h: & h - \text{constant} \\ \kappa: & \kappa \ge 0 \\ sign: & J_j g_j \ge 0, \quad j = 1,\ldots,N \\ \ell_0: & \lim_{\varepsilon \to 0} \sum_{j=1}^{N} \left|J_j\right|^\varepsilon = Np \\ gap: & \left|J_j\right| \ge \Delta_j \ge 0 \quad if \quad J_j \ne 0 \end{cases} \tag{A2}$$

Next, building on the methodology of Bouten et al., [11], a connection strength $J_j$ was decomposed into a product of three independent variables: sign $s_j$ (+1 if positive and –1 if negative), connectedness $c_j$ (0 if connection strength is zero and 1 otherwise), and a non-negative connection weight $w_j$.

$$J_j = s_j c_j w_j, \quad s_j \in \{-1,1\}, \quad c_j \in \{0,1\}, \quad w_j \geq 0 \tag{A3}$$

This substitution made it possible to linearize the absolute value function appearing in the $\ell_1$ norm and *gap* constraints and simplified the $\ell_0$ norm constraint, leading to the following equations.

$$\left(2y^\mu - 1\right)\left(\sum_{j=1}^{N} s_j c_j w_j X_j^\mu - h\right) > \kappa, \quad \mu = 1,\ldots,m$$

$$\text{Prob}\left(X_j^\mu\right) = \begin{cases} 1-f_j, & X_j^\mu = 0 \\ f_j, & X_j^\mu = 1 \end{cases}; \quad \text{Prob}\left(y^\mu\right) = \begin{cases} 1-f_{out}, & y^\mu = 0 \\ f_{out}, & y^\mu = 1 \end{cases}$$

subject to:

$$\begin{cases} \ell_1: & \sum_{j=1}^{N} c_j w_j = Nw \\ h: & h - \text{constant} \\ \kappa: & \kappa \geq 0 \\ sign: & s_j g_j > 0, \quad j = 1,\ldots,N \\ \ell_0: & \sum_{j=1}^{N} c_j = Np \\ gap: & w_j \geq \Delta_j c_j \geq 0, \quad j = 1,\ldots,N \end{cases} \tag{A4}$$

The objective of this model is to store a set of $m$ associations $\left\{X^\mu \to y^\mu\right\}$ by learning the variables $\{s_j\},\{c_j\},\{w_j\}$, given the fixed parameters $N, h, \kappa, p, w, \{g_j\}, \{\Delta_j\}, \{f_j\}, f_{out}$.

The $\ell_1$ norm constraint ensures that the solution volume of Eqs. (A2) is finite in the space of connection strengths $\{J_j\}$. However, in the space of variables $\{s_j, c_j, w_j\}$ used in Eqs. (A4), the solution volume can be infinite because a connection with $c_j = 0$ can have an arbitrary weight $w_j \geq 0$. To keep this engineered divergence from getting in the way of replica calculations, we used an additional regularizing constraint on connections with $c_j = 0$. As this constraint is introduced for convenience only, its exact form is not expected to affect the observables of the original model, Eqs. (A2). We used $\sum_{j=1}^{N}\left(1-c_j\right)w_j = Nw_x$ with the expectation that the parameter $w_x$ would not appear in the results.

Additional assumptions about the scaling of model parameters with $N$ are needed to take the $N \to \infty$ limit required by the replica method. For this, we used a previously established associative learning scaling convention [12,13]. We assumed that $m/N$, $p$, $f_j$, and $f_{out}$ are of order 1 in $N$. We also defined scaling for the remaining model parameters and variables by introducing the following normalized notation.

$$\frac{w_j}{w} = \tilde{w}_j; \quad \frac{\Delta_j}{w} = \tilde{\Delta}_j; \quad \frac{h}{w} = N\tilde{h}; \quad \frac{w_x}{w} = \tilde{w}_x; \quad \frac{\kappa}{w} = \sqrt{N}\tilde{\kappa} \tag{A5}$$

The normalized quantities in these expressions (indicated with tilde), do not scale with $N$. We note that in the case of a learnable threshold (or no threshold, $h$ = 0), the details of scaling become unimportant [16], and another convention such as the one used in balanced networks [15,28] would lead to identical results. With Eqs. (A5), the model transformed into,

$$\left(2y^{\mu}-1\right)\left(\frac{1}{N}\sum_{j=1}^{N}s_j c_j \tilde{w}_j X_j^{\mu}-\tilde{h}\right)>\frac{\tilde{\kappa}}{\sqrt{N}},\quad \mu=1,\ldots,m$$

$$\text{Prob}\left(X_j^{\mu}\right)=\begin{cases}1-f_j, & X_j^{\mu}=0\\ f_j, & X_j^{\mu}=1\end{cases};\quad \text{Prob}\left(y^{\mu}\right)=\begin{cases}1-f_{out}, & y^{\mu}=0\\ f_{out}, & y^{\mu}=1\end{cases}$$

subject to:

$$\begin{cases}\ell_1: \quad \sum_{j=1}^{N}c_j\tilde{w}_j=N\\ h: \quad \tilde{h}-\text{constant}\\ \kappa: \quad \tilde{\kappa}\geq 0\\ sign: \quad s_j g_j>0, \quad j=1,\ldots,N\\ \ell_0: \quad \sum_{j=1}^{N}c_j=Np\\ gap: \quad \tilde{w}_j\geq\tilde{\Delta}_j c_j\geq 0, \quad j=1,\ldots,N\\ \sum_{j=1}^{N}\left(1-c_j\right)\tilde{w}_j=N\tilde{w}_x\end{cases} \tag{A6}$$

**Mathematical preliminaries**

The following notation, identities, and approximations were used in the replica calculations below.

Shorthand notation: $\left(X\right)_{a>b}=X\theta\left(a-b\right)$ and $\left(X\right)_{a<b}=X\theta\left(b-a\right)$, where $\theta$ is the Heaviside step-function.

Fourier representations of Dirac delta-function, Heaviside step-function, and Kronecker delta:

$$\delta\left(x\right)=\int\frac{d\hat{x}}{2\pi}e^{i\hat{x}x};\quad \theta\left(x\right)=\int\frac{d'ud\hat{u}}{2\pi}e^{i\hat{u}\left(x-u\right)};\quad \delta_{x,x_0}=\int\frac{d''\hat{\varphi}}{2\pi}e^{i\hat{\varphi}\left(x-x_0\right)} \tag{A7}$$

Symbol $d$ in these expressions and thereafter was used to indicate integration from $-\infty$ to $\infty$, $d'$ for 0 to $\infty$, and $d''$ for 0 to $2\pi$.

Hubbard-Stratonovich transformation:

$$e^{a\sum_{i,j=1}^{N} s_i s_j} = \int_{-\infty}^{\infty} \frac{dx}{\sqrt{\pi}} e^{-x^2 + 2x\sqrt{a}\sum_{i=1}^{N} s_i} \tag{A8}$$

Useful integrals:

$$\begin{aligned}
&\int_{-\infty}^{\infty} \frac{dx}{\sqrt{\pi}} e^{-ax^2+bx} = \frac{1}{\sqrt{a}} e^{\frac{b^2}{4a}}, \quad a > 0 \\
&\int_{\Delta}^{\infty} \frac{dx}{\sqrt{\pi}} e^{-ax^2+bx} = \frac{1}{2\sqrt{a}} e^{\frac{b^2}{4a}} \operatorname{erfc}\left( \frac{-b}{2\sqrt{a}} + \sqrt{a}\Delta \right), \quad a > 0
\end{aligned} \tag{A9}$$

Special functions $B$, $D$, $E$, $F$, and identities involving these functions:

$$\begin{aligned}
&B(x) = \frac{e^{-x^2}}{\sqrt{\pi}}; \; E(x) = \frac{1}{2}\big(1 + \operatorname{erf}(x)\big); \; F(x) = B(x) + 2xE(x); \; D(x) = xF(x) + E(x) \\
&\frac{d}{dx} E(x) = B(x); \quad \frac{d}{dx} F(x) = 2E(x); \quad \frac{d}{dx} D(x) = 2F(x) \\
&\int_{z}^{\infty} \frac{dx}{\sqrt{\pi}} e^{-x^2} = E(-z); \quad \int_{z}^{\infty} \frac{dx}{\sqrt{\pi}} e^{-x^2} (x - z) = \frac{1}{2} F(-z); \quad \int_{z}^{\infty} \frac{dx}{\sqrt{\pi}} e^{-x^2} (x - z)^2 = \frac{1}{2} D(-z)
\end{aligned}$$

(A10)

Asymptotic expansions involving the complementary error function:

$$\operatorname{erfc}(x)=\begin{cases}\dfrac{e^{-x^2}}{x\sqrt{\pi}}\left(1+O\left(x^{-2}\right)\right), & x\to+\infty\\ 2+\dfrac{e^{-x^2}}{x\sqrt{\pi}}\left(1+O\left(x^{-2}\right)\right), & x\to-\infty\end{cases}\ ;\ \ln\left(\operatorname{erfc}(x)\right)=\begin{cases}-x^2+O(\ln x), & x\to+\infty\\ \ln 2+O\left(\dfrac{e^{-x^2}}{x}\right), & x\to-\infty\end{cases}$$

$$\int_{-\infty}^{\infty}\frac{e^{-x^2}}{\sqrt{\pi}}dx\ln\left(\operatorname{erfc}\left(\frac{a-x}{\sqrt{\varepsilon}}\right)\right)=\frac{-D(a)}{2\varepsilon}+O(\ln\varepsilon),\quad \varepsilon\to+0$$

$$\frac{a}{a+\sqrt{\varepsilon}e^{\frac{1}{\delta}\left(x^2-c^2\right)}\operatorname{erfc}\left(\frac{1}{\sqrt{\varepsilon}}(-x+b)\right)}\approx\theta\left(-x+\frac{c^2+\max(b,c)^2}{2\max(b,c)}\right),\quad b,c>0$$

$$\ln\left(a+\sqrt{\varepsilon}e^{\frac{1}{\varepsilon}\left(x^2-c^2\right)}\operatorname{erfc}\left(\frac{1}{\sqrt{\varepsilon}}(-x+b)\right)\right)_{c\geq b}\approx\begin{cases}\dfrac{1}{\varepsilon}\left(x^2-c^2\right), & x>c\\ \ln a, & x<c\end{cases}$$

$$\ln\left(a+\sqrt{\varepsilon}e^{\frac{1}{\varepsilon}\left(x^2-c^2\right)}\operatorname{erfc}\left(\frac{1}{\sqrt{\varepsilon}}(-x+b)\right)\right)_{c<b}\approx\begin{cases}\dfrac{1}{\varepsilon}\left(x^2-c^2\right), & x>b\\ \dfrac{2b}{\varepsilon}\left(x-\dfrac{b^2+c^2}{2b}\right), & \dfrac{b^2+c^2}{2b}<x<b\\ \ln a, & x<\dfrac{b^2+c^2}{2b}\end{cases}\tag{A11}$$

**Replica solution of the model**

We solved the constrained learning model of Eqs. (A6) by extending the previously described replica calculations [10-12,16] to work with a combination of continuous variables $\{w_j\}$ and binary variables $\{s_j,c_j\}$. To that end, we first calculated the solution space volume of Eqs. (A6) for a given set of associations,

$$\Omega\left(\left\{X_j^\mu\right\},\left\{y^\mu\right\}\right)=\sum_{\{s_j,c_j\}}\int\prod_{j=1}^{N}d'\tilde{w}_j\prod_{\mu=1}^{m}\theta\left(\left(2y^\mu-1\right)\left(\frac{1}{N}\sum_{j=1}^{N}s_jc_j\tilde{w}_jX_j^\mu-\tilde{h}\right)-\frac{\tilde{\kappa}}{\sqrt{N}}\right)\times$$
$$\delta\left(\sum_{j=1}^{N}c_j\tilde{w}_j-N\right)\delta_{\sum_{j=1}^{N}c_j,Np}\prod_{j=1}^{N}\theta\left(\tilde{w}_j-\tilde{\Delta}_jc_j\right)\delta\left(\sum_{j=1}^{N}\left(1-c_j\right)\tilde{w}_j-N\tilde{w}_x\right)\tag{A12}$$

This expression incorporates a fixed threshold, robustness, and four constraints on connectivity. By solving this general problem, it is possible to reduce the solution to an arbitrary combination of these features ($2^6$ cases in total). When *sign* constraints are present, $s_j = g_j$, and the summation over $s_j$ in Eq. (A12) must be dropped.

We defined the typical volume of the solution space, $\Omega_{typical}$, through the average of $\ln\left(\Omega\left(X_j^\mu, y^\mu\right)\right)$ over the associations, and calculated this average by introducing $n$ replica systems,

$$\ln\left(\Omega_{typical}\right) = \left\langle \ln\left(\Omega\left(\left\{X_j^\mu\right\},\left\{y^\mu\right\}\right)\right)\right\rangle_{\left\{X_j^\mu\right\},\left\{y^\mu\right\}} = \lim_{n\to 0} \frac{\left\langle \Omega\left(\left\{X_j^\mu\right\},\left\{y^\mu\right\}\right)^n \right\rangle_{\left\{X_j^\mu\right\},\left\{y^\mu\right\}} - 1}{n} \qquad \text{(A13)}$$

The product of $n$ replica volumes (enumerated below with a superscript $a$) was then written as a single multidimensional integral.

$$\left\langle \Omega\left(\left\{X_j^\mu\right\},\left\{y^\mu\right\}\right)^n \right\rangle_{\left\{X_j^\mu\right\},\left\{y^\mu\right\}} = \sum_{\left\{s_j^a, c_j^a\right\}} \left[ \int \prod_{a,j=1}^{n,N} d'\tilde{w}_j^a \times \right.$$
$$\left\langle \prod_{a,\mu=1}^{n,m} \theta\left(\left(2y^\mu - 1\right)\left(\frac{1}{N}\sum_{j=1}^{N} s_j^a c_j^a \tilde{w}_j^a X_j^\mu - \tilde{h}\right) - \frac{\tilde{\kappa}}{\sqrt{N}}\right)\right\rangle_{\left\{X_j^\mu\right\},\left\{y^\mu\right\}} \times \qquad \text{(A14)}$$
$$\left. \prod_{a=1}^{n} \delta\left(\sum_{j=1}^{N} c_j^a \tilde{w}_j^a - N\right) \prod_{a=1}^{n} \delta_{\sum_{j=1}^{N} c_j^a, Np} \prod_{a,j=1}^{n,N} \theta\left(\tilde{w}_j^a - \tilde{\Delta}_j c_j^a\right) \prod_{a=1}^{n} \delta\left(\sum_{j=1}^{N}\left(1 - c_j^a\right)\tilde{w}_j^a - N\tilde{w}_x\right)\right]$$

New variables, $\frac{\lambda^{a,\mu}}{\sqrt{N}} = \frac{1}{N}\sum_{j=1}^{N} s_j^a c_j^a \tilde{w}_j^a X_j^\mu - \tilde{h}$, were introduced to decouple the input and output associations, $X_j^\mu$ and $y^\mu$, and facilitate the averaging.

$$\left\langle \Omega\left(\left\{X_j^\mu\right\},\left\{y^\mu\right\}\right)^n \right\rangle_{\left\{X_j^\mu\right\},\left\{y^\mu\right\}} = \sum_{\left\{s_j^a c_j^a\right\}} \left[ \int \prod_{a,j=1}^{n,N} d'\tilde{w}_j^a \prod_{\mu,a=1}^{m,n} \frac{d\lambda^{a,\mu}}{\sqrt{N}} \times \right.$$
$$\left\langle \prod_{a,\mu=1}^{n,m} \theta\left(\left(2y^\mu - 1\right)\frac{\lambda^{a,\mu}}{\sqrt{N}} - \frac{\tilde{\kappa}}{\sqrt{N}}\right) \right\rangle_{\left\{y^\mu\right\}} \left\langle \prod_{\mu,a=1}^{m,n} \delta\left(\frac{\lambda^{a,\mu}}{\sqrt{N}} - \frac{1}{N}\sum_{j=1}^{N} s_j^a c_j^a \tilde{w}_j^a X_j^\mu + \tilde{h}\right) \right\rangle_{\left\{X_j^\mu\right\}} \times \quad \text{(A15)}$$
$$\left. \prod_{a=1}^{n} \delta\left(\sum_{j=1}^{N} c_j^a \tilde{w}_j^a - N\right) \prod_{a=1}^{n} \delta_{\sum_{j=1}^{N} c_j^a, Np} \prod_{a,j=1}^{n,N} \theta\left(\tilde{w}_j^a - \tilde{\Delta}_j c_j^a\right) \prod_{a=1}^{n} \delta\left(\sum_{j=1}^{N}\left(1 - c_j^a\right)\tilde{w}_j^a - N\tilde{w}_x\right) \right]$$

The delta and theta functions were replaced with their integral representations, Eqs. (A7), and the averaging over $X_j^\mu$ and $y^\mu$ was performed.

$$\left\langle \Omega\left(\left\{X_j^\mu\right\},\left\{y^\mu\right\}\right)^n \right\rangle_{\left\{X_j^\mu\right\},\left\{y^\mu\right\}} = \sum_{\left\{s_j^a, c_j^a\right\}} \left[ \int \prod_{a,j=1}^{n,N} \left(d'\tilde{w}_j^a \theta\left(\tilde{w}_j^a - \tilde{\Delta}_j c_j^a\right)\right) \prod_{\mu,a=1}^{m,n} \frac{d\lambda^{a,\mu} d\hat{\lambda}^{a,\mu}}{2\pi} \times \right.$$
$$\prod_{a,\mu=1}^{n,m} \frac{d'u^{a,\mu} d\hat{u}^{a,\mu}}{2\pi} \prod_{a=1}^{n} \frac{d''\hat{\varphi}^a}{2\pi} \prod_{a=1}^{n} \frac{d\hat{k}^a}{2\pi} \prod_{a=1}^{n} \frac{d\hat{v}^a}{2\pi} e^{-i\sum_{a,\mu=1}^{n,m} \hat{u}^{a,\mu}\left(\tilde{\kappa} + u^{a,\mu}\right)} \times$$
$$\prod_{\mu=1}^{m} \left( f_{out} e^{i\sum_{a=1}^{n} \lambda^{a,\mu}\hat{u}^{a,\mu}} + \left(1 - f_{out}\right) e^{-i\sum_{a=1}^{n} \lambda^{a,\mu}\hat{u}^{a,\mu}} \right) e^{\sum_{a,\mu=1}^{n,m} i\hat{\lambda}^{a,\mu}\left(\lambda^{a,\mu} + \sqrt{N}\tilde{h}\right)} \times \quad \text{(A16)}$$
$$\left. \prod_{j,\mu=1}^{N,m} \left( 1 - f_j + f_j e^{-\frac{i}{\sqrt{N}}\sum_{a=1}^{n} \hat{\lambda}^{a,\mu} s_j^a c_j^a \tilde{w}_j^a} \right) e^{\sum_{a=1}^{n} i\hat{\varphi}^a\left(\sum_{j=1}^{N} c_j^a - Np\right)} e^{\sum_{a=1}^{n} i\hat{k}^a\left(\sum_{j=1}^{N} c_j^a \tilde{w}_j^a - N\right)} e^{\sum_{a=1}^{n} i\hat{v}^a\left(\sum_{j=1}^{N}\left(1 - c_j^a\right)\tilde{w}_j^a - N\tilde{w}_x\right)} \right]$$

Replacing the argument of the first product in line four of Eq. (A16) with an exponential expression that approximates it up to the second order in $\frac{-i}{\sqrt{N}}\sum_{a=1}^{n} \hat{\lambda}^{a,\mu} s_j^a c_j^a \tilde{w}_j^a$, we obtained the following.

$$\left\langle \Omega\left(\left\{X_j^\mu\right\},\left\{y^\mu\right\}\right)^n \right\rangle_{\left\{X_j^\mu\right\},\left\{y^\mu\right\}} = \sum_{\left\{s_j^a,c_j^a\right\}} \Bigg[ \int \prod_{a,j=1}^{n,N} \left( d'\tilde{w}_j^a \theta\left(\tilde{w}_j^a - \tilde{\Delta}_j c_j^a\right)\right) \prod_{\mu,a=1}^{m,n} \frac{d\lambda^{a,\mu} d\hat{\lambda}^{a,\mu}}{2\pi} \times$$

$$\prod_{a,\mu=1}^{n,m} \frac{d'u^{a,\mu} d\hat{u}^{a,\mu}}{2\pi} \prod_{a=1}^{n} \frac{d''\hat{\varphi}^a}{2\pi} \prod_{a=1}^{n} \frac{d\hat{k}^a}{2\pi} \prod_{a=1}^{n} \frac{d\hat{v}^a}{2\pi} e^{-i\sum_{a,\mu=1}^{n,m} \hat{u}^{a,\mu}\left(\tilde{\kappa}+u^{a,\mu}\right)} \times$$

$$\prod_{\mu=1}^{m} \left( f_{out} e^{i\sum_{a=1}^{n}\lambda^{a,\mu}\hat{u}^{a,\mu}} + \left(1-f_{out}\right) e^{-i\sum_{a=1}^{n}\lambda^{a,\mu}\hat{u}^{a,\mu}} \right) e^{\sum_{a,\mu=1}^{n,m} i\hat{\lambda}^{a,\mu}\left(\lambda^{a,\mu}+\sqrt{N}\tilde{h}\right)} \times \tag{A17}$$

$$\prod_{\mu=1}^{m} e^{-i\sum_{a=1}^{n}\hat{\lambda}^{a,\mu}\left(\frac{1}{\sqrt{N}}\sum_{j=1}^{N} f_j s_j^a c_j^a \tilde{w}_j^a\right) - \sum_{a,b=1}^{n,n}\hat{\lambda}^{a,\mu}\hat{\lambda}^{b,\mu}\left(\frac{1}{2N}\sum_{j=1}^{N} f_j\left(1-f_j\right) s_j^a s_j^b c_j^a c_j^b \tilde{w}_j^a \tilde{w}_j^b\right)} \times$$

$$e^{\sum_{a=1}^{n} i\hat{\varphi}^a\left(\sum_{j=1}^{N} c_j^a - Np\right)} e^{\sum_{a=1}^{n} i\hat{k}^a\left(\sum_{j=1}^{N} c_j^a \tilde{w}_j^a - N\right)} e^{\sum_{a=1}^{n} i\hat{v}^a\left(\sum_{j=1}^{N}\left(1-c_j^a\right)\tilde{w}_j^a - N\tilde{w}_x\right)} \Bigg]$$

Two sets of order parameters were introduced and inserted into Eq. (A17) to decouple the products containing indices $j$ and $\mu$,

$$\frac{1}{\sqrt{N}}\sum_{j=1}^{N} f_j s_j^a c_j^a \tilde{w}_j^a = \sqrt{N}\tilde{h} + s^a, \quad \frac{1}{N}\sum_{j=1}^{N} f_j\left(1-f_j\right) s_j^a s_j^b c_j^a c_j^b \tilde{w}_j^a \tilde{w}_j^b = q^{a,b} \tag{A18}$$

The result was expressed in a standard form, using functions $\tilde{G}_E$ and $\tilde{G}_S$:

$$\left\langle \Omega\left(\left\{X_j^\mu\right\},\left\{y^\mu\right\}\right)^n \right\rangle_{\left\{X_j^\mu\right\},\left\{y^\mu\right\}} = \int \prod_{a=1}^{n} \frac{d''\hat{\varphi}^a}{2\pi} \prod_{a=1}^{n} \frac{d\hat{k}^a}{2\pi} \prod_{a=1}^{n} \frac{d\hat{v}^a}{2\pi} \prod_{a=1}^{n} \frac{ds^a d\hat{s}^a}{2\pi/\sqrt{N}} \prod_{a,b=1}^{n,n} \frac{dq^{a,b} d\hat{q}^{a,b}}{2\pi/N} \times$$

$$e^{N\left(i\sum_{a,b=1}^{n,n} q^{a,b}\hat{q}^{a,b} - ip\sum_{a=1}^{n}\hat{\varphi}^a + i\tilde{h}\sum_{a=1}^{n}\hat{s}^a - i\sum_{a=1}^{n}\left(\hat{k}^a + \hat{v}^a\tilde{w}_x\right) + n\alpha\tilde{G}_E\left(\left\{s^a\right\},\left\{q^{a,b}\right\}\right) + n\tilde{G}_S\left(\left\{\hat{\varphi}^a\right\},\left\{\hat{k}^a\right\},\left\{\hat{v}^a\right\},\left\{\hat{s}^a\right\},\left\{\hat{q}^{a,b}\right\}\right)\right)}$$

$$\tilde{G}_E\left(\left\{s^a\right\},\left\{q^{a,b}\right\}\right) = \frac{1}{n}\ln\left[\int \prod_{a=1}^{n} \frac{d'u^a d\hat{u}^a}{2\pi} e^{-\frac{1}{2}\sum_{a,b=1}^{n}\hat{u}^a\hat{u}^b q^{a,b}} \times \right.$$

$$\left.\left( f_{out} e^{i\sum_{a=1}^{n}\hat{u}^a\left(s^a - u^a - \tilde{\kappa}\right)} + (1-f_{out}) e^{-i\sum_{a=1}^{n}\hat{u}^a\left(s^a + u^a + \tilde{\kappa}\right)} \right)\right]$$

$$\tilde{G}_S\left(\left\{\hat{\varphi}^a\right\},\left\{\hat{k}^a\right\},\left\{\hat{v}^a\right\},\left\{\hat{s}^a\right\},\left\{\hat{q}^{a,b}\right\}\right) = \frac{1}{n}\frac{1}{N}\ln \sum_{\left\{s_j^a, c_j^a\right\}} \prod_{j=1}^{N} \left[\int \prod_{a=1}^{n}\left(d'\tilde{w}_j^a \theta\left(\tilde{w}_j^a - \tilde{\Delta}_j c_j^a\right)\right) \times \right.$$

$$\left. e^{i\sum_{a=1}^{n}\left(-f_j\hat{s}^a s_j^a c_j^a \tilde{w}_j^a + \hat{\varphi}^a c_j^a + \hat{k}^a c_j^a \tilde{w}_j^a + \hat{v}^a\left(1-c_j^a\right)\tilde{w}_j^a\right) - if_j\left(1-f_j\right)\sum_{a,b=1}^{n,n}\hat{q}^{a,b} s_j^a s_j^b c_j^a c_j^b \tilde{w}_j^a \tilde{w}_j^b} \right] \tag{A19}$$

The integral in the first line of Eqs. (A19) was calculated by using the steepest descent method combined with the assumption of a replica-symmetric saddle point: $\varphi^a = \varphi$, $\hat{k}^a = \hat{k}$, $\hat{v}^a = \hat{v}$, $s^a = s$, $\hat{s}^a = \hat{s}$, $q^{a,a} = q_0$, $q^{a \neq b} = q$, $\hat{q}^{a,a} = \hat{q}_0$, and $\hat{q}^{a \neq b} = \hat{q}$.

$$\left\langle \Omega\left(\left\{X_j^\mu\right\},\left\{y^\mu\right\}\right)^n \right\rangle_{\left\{X_j^\mu\right\},\left\{y^\mu\right\}} = e^{Nn\left(-ip\hat{\varphi} + i\tilde{h}\hat{s} - i\hat{k} - i\hat{v}\tilde{w}_x + iq_0\hat{q}_0 - iq\hat{q} + \alpha G_E\left(s,q_0,q\right) + G_S\left(\hat{\varphi},\hat{k},\hat{v},\hat{s},\hat{q}_0,\hat{q}\right)\right)}$$

$$G_E\left(s,q_0,q\right) = \frac{1}{n}\ln\left[\int \prod_{a=1}^{n} \frac{d'u^a d\hat{u}^a}{2\pi} e^{-\frac{1}{2}q_0\sum_{a=1}^{n}\hat{u}^a\hat{u}^a - \frac{1}{2}q\sum_{a\neq b=1}^{n,n}\hat{u}^a\hat{u}^b} \times \right.$$

$$\left.\left( f_{out} e^{i\sum_{a=1}^{n}\hat{u}^a\left(s - u^a - \tilde{\kappa}\right)} + (1-f_{out}) e^{-i\sum_{a=1}^{n}\hat{u}^a\left(s + u^a + \tilde{\kappa}\right)} \right)\right] \tag{A20}$$

$$G_S\left(\hat{\varphi},\hat{k},\hat{v},\hat{s},\hat{q}_0,\hat{q}\right) = \frac{1}{n}\frac{1}{N}\ln \sum_{\left\{s_j^a, c_j^a\right\}} \prod_{j=1}^{N} \left[\int \prod_{a=1}^{n}\left(d'\tilde{w}_j^a \theta\left(\tilde{w}_j^a - \tilde{\Delta}_j c_j^a\right)\right) \times \right.$$

$$\left. e^{i\sum_{a=1}^{n}\left(-f_j\hat{s}s_j^a c_j^a \tilde{w}_j^a + \hat{\varphi}c_j^a + \hat{k}c_j^a \tilde{w}_j^a + \hat{v}\left(1-c_j^a\right)\tilde{w}_j^a\right) - if_j\left(1-f_j\right)\hat{q}_0\sum_{a=1}^{n} s_j^a s_j^a c_j^a c_j^a \tilde{w}_j^a \tilde{w}_j^a - if_j\left(1-f_j\right)\hat{q}\sum_{a\neq b=1}^{n,n} s_j^a s_j^b c_j^a c_j^b \tilde{w}_j^a \tilde{w}_j^b} \right]$$

The replica-symmetric saddle point coordinates $\left(s,q_0,q,\hat{\varphi},\hat{k},\hat{v},\hat{s},\hat{q}_0,\hat{q}\right)$ satisfy the following system of equations:

$$
\begin{aligned}
&\frac{\partial G_E\left(s,q_0,q\right)}{\partial s}=0;\quad i\hat{q}_0+\alpha\frac{\partial G_E\left(s,q_0,q\right)}{\partial q_0}=0;\quad -i\hat{q}+\alpha\frac{\partial G_E\left(s,q_0,q\right)}{\partial q}=0\\
&-ip+\frac{\partial G_S\left(\hat{\varphi},\hat{k},\hat{v},\hat{s},\hat{q}_0,\hat{q}\right)}{\partial\hat{\varphi}}=0;\quad -i+\frac{\partial G_S\left(\hat{\varphi},\hat{k},\hat{v},\hat{s},\hat{q}_0,\hat{q}\right)}{\partial\hat{k}}=0;\\
&-i\tilde{w}_x+\frac{\partial G_S\left(\hat{\varphi},\hat{k},\hat{v},\hat{s},\hat{q}_0,\hat{q}\right)}{\partial\hat{v}}=0;\quad i\tilde{h}+\frac{\partial G_S\left(\hat{\varphi},\hat{k},\hat{v},\hat{s},\hat{q}_0,\hat{q}\right)}{\partial\hat{s}}=0;\\
&iq_0+\frac{\partial G_S\left(\hat{\varphi},\hat{k},\hat{v},\hat{s},\hat{q}_0,\hat{q}\right)}{\partial\hat{q}_0}=0;\quad -iq+\frac{\partial G_S\left(\hat{\varphi},\hat{k},\hat{v},\hat{s},\hat{q}_0,\hat{q}\right)}{\partial\hat{q}}=0
\end{aligned}
\tag{A21}
$$

Functions $G_E$ and $G_S$ were simplified after applying the Hubbard-Stratonovich transformation, Eq. (A8), and taking the $n\to 0$ limit.

$$
\begin{aligned}
&G_E\left(s,q_0,q\right)=-\ln 2+\int_{-\infty}^{\infty}\frac{e^{-x^2}}{\sqrt{\pi}}dx\times\\
&\left(f_{out}\ln\left(\operatorname{erfc}\left(\frac{-s+\tilde{\kappa}-x\sqrt{2q}}{\sqrt{2\left(q_0-q\right)}}\right)\right)+\left(1-f_{out}\right)\ln\left(\operatorname{erfc}\left(\frac{s+\tilde{\kappa}-x\sqrt{2q}}{\sqrt{2\left(q_0-q\right)}}\right)\right)\right)\\
&G_S\left(\hat{\varphi},\hat{k},\hat{v},\hat{s},\hat{q}_0,\hat{q}\right)=\frac{1}{N}\sum_{j=1}^{N}\int_{-\infty}^{\infty}\frac{e^{-x^2}}{\sqrt{\pi}}dx\ln\sum_{s_j}\left[e^{i\hat{\varphi}}\frac{\sqrt{\pi}}{2}\frac{e^{\left(-xs_j\sqrt{\frac{-\hat{q}}{\hat{q}_0-\hat{q}}}+\frac{if_js_j\hat{s}-i\hat{k}}{2\sqrt{if_j\left(1-f_j\right)\left(\hat{q}_0-\hat{q}\right)}}\right)^2}}{\sqrt{if_j\left(1-f_j\right)\left(\hat{q}_0-\hat{q}\right)}}\times\right.\\
&\left.\operatorname{erfc}\left(-xs_j\sqrt{\frac{-\hat{q}}{\hat{q}_0-\hat{q}}}+\frac{if_js_j\hat{s}-i\hat{k}}{2\sqrt{if_j\left(1-f_j\right)\left(\hat{q}_0-\hat{q}\right)}}+\sqrt{if_j\left(1-f_j\right)\left(\hat{q}_0-\hat{q}\right)}\tilde{\Delta}_j\right)+\frac{i}{\hat{v}}\right]
\end{aligned}
\tag{A22}
$$

We introduced new variables, $u_{\pm}=\frac{\tilde{\kappa}\pm s}{\sqrt{2q}}$, $\varepsilon=\frac{q_0-q}{q}$, $t=-i\hat{q}$, $z=\frac{i\hat{s}}{2\sqrt{-i\hat{q}}}$, $\delta=-\frac{\hat{q}_0-\hat{q}}{\hat{q}}$, $\eta=-\frac{i\hat{k}}{\sqrt{-i\hat{q}}}$, $\nu=\frac{i}{\hat{v}}$, and $\varphi=-i\hat{\varphi}$, and rewrote Eqs. (A21, A22) in terms of only real quantities.

$$G_E\left(u_+,u_-,\varepsilon\right)=-\ln 2+\int_{-\infty}^{\infty}\frac{e^{-x^2}}{\sqrt{\pi}}dx\left(f_{out}\ln\left(\operatorname{erfc}\left(\frac{u_- - x}{\sqrt{\varepsilon}}\right)\right)+\left(1-f_{out}\right)\ln\left(\operatorname{erfc}\left(\frac{u_+ - x}{\sqrt{\varepsilon}}\right)\right)\right)$$

$$G_S\left(\varphi,\nu,\eta,t,z,\delta\right)=\frac{1}{N}\sum_{j=1}^{N}\int_{-\infty}^{\infty}\frac{e^{-x^2}}{\sqrt{\pi}}dx\ln\sum_{s_j}\left[\nu+e^{-\varphi}\sqrt{\frac{\pi}{4f_j\left(1-f_j\right)\delta t}}e^{\frac{1}{\delta}\left(-xs_j+\frac{2f_js_jz+\eta}{\sqrt{4f_j\left(1-f_j\right)}}\right)^2}\times\right. \quad \text{(A23)}$$

$$\left.\operatorname{erfc}\left(\frac{1}{\sqrt{\delta}}\left(-xs_j+\frac{2f_js_jz+\eta}{\sqrt{4f_j\left(1-f_j\right)}}+\sqrt{f_j\left(1-f_j\right)t\delta^2}\tilde{\Delta}_j\right)\right)\right]$$

The saddle-point equations, Eqs. (A21), were transformed as follows.

$$\begin{cases}\dfrac{\partial G_E\left(u_+,u_-,\varepsilon\right)}{\partial u_+}-\dfrac{\partial G_E\left(u_+,u_-,\varepsilon\right)}{\partial u_-}=0\\[2mm] \dfrac{\partial G_E\left(u_+,u_-,\varepsilon\right)}{\partial \varepsilon}=\dfrac{t}{\alpha}\dfrac{2\tilde{\kappa}^2}{\left(u_+ + u_-\right)^2}\\[2mm] \dfrac{\partial G_E\left(u_+,u_-,\varepsilon\right)}{\partial u_+}=\dfrac{4\tilde{\kappa}^2 t}{\alpha\left(u_+ + u_-\right)^3}\left(\delta-\varepsilon\right)\\[2mm] \dfrac{\partial G_S\left(\varphi,\nu,\eta,t,z,\delta\right)}{\partial \varphi}=-p\\[2mm] \dfrac{\partial G_S\left(\varphi,\nu,\eta,t,z,\delta\right)}{\partial \eta}=-\sqrt{t}\\[2mm] \dfrac{\partial G_S\left(\varphi,\nu,\eta,t,z,\delta\right)}{\partial \nu}=\dfrac{\tilde{w}_x}{\nu^2}\\[2mm] \dfrac{\partial G_S\left(\varphi,\nu,\eta,t,z,\delta\right)}{\partial z}=-2\sqrt{t}\tilde{h}\\[2mm] \dfrac{\partial G_S\left(\varphi,\nu,\eta,t,z,\delta\right)}{\partial \delta}=-t\dfrac{2\tilde{\kappa}^2}{\left(u_+ + u_-\right)^2}\\[2mm] \dfrac{\partial G_S\left(\varphi,\nu,\eta,t,z,\delta\right)}{\partial t}=-\dfrac{1}{2\sqrt{t}}\left(2\tilde{h}z+\eta\right)-\dfrac{2\tilde{\kappa}^2}{\left(u_+ + u_-\right)^2}\left(\delta-\varepsilon\right)\end{cases} \quad \text{(A24)}$$

Given the memory load, $\alpha$, Eqs. (A23, A24) can be solved for the 9 saddle-point variables, making it possible to find the typical volume of the solution region.

$$\frac{\ln\left(\Omega_{typical}\right)}{N} = p\varphi + \left(2\tilde{h}z + \eta\right)\sqrt{t} + \frac{\tilde{x}}{\nu} + \frac{2\tilde{\kappa}^2 t\left(\delta - \varepsilon\right)}{\left(u_+ + u_-\right)^2} + \alpha G_E\left(u_+, u_-, \varepsilon\right) + G_S\left(\varphi, \nu, \eta, t, z, \delta\right)$$

(A25)

We note two obvious constraints on saddle-point variables. First, the assumption of a replica-symmetric saddle point, coupled with the fact that $q_0$ and $q$ contain within- and between-replica products respectively [see Eqs. (A18)], implies that $q_0 \geq q$, and thus, $\varepsilon \geq 0$. Second, as $\tilde{\kappa} \geq 0$, $u_+ + u_- = \frac{2\tilde{\kappa}}{\sqrt{2q}} \geq 0$.

## Replica solution at critical capacity

At critical associative memory storage capacity, $\alpha_c$, the saddle point Eqs. (A24) can be simplified because as $\Omega_{typical}$ tends to zero in the $\alpha \to \alpha_c$ limit, $\varepsilon = \frac{q_0 - q}{q}$ goes to zero as well. In this limit, functions $G_E$ and $G_S$ can be expanded asymptotically up to the leading orders in $\varepsilon^{-1}$ and $\delta^{-1}$ using Eqs. (A10, A11).

$$G_E\left(u_+, u_-, \varepsilon\right) = \frac{-1}{2\varepsilon}\left(f_{out} D\left(u_-\right) + \left(1 - f_{out}\right) D\left(u_+\right)\right)$$

$$G_S\left(\varphi, \nu, \eta, t, z, \delta\right) = \frac{1}{N}\sum_{j=1}^{N}\left(\begin{array}{l}\left(1 - \sum_{s_j} E\left(-a_j - \sqrt{\delta\varphi}\right)\right)\ln\sum_{s_j}\nu + \\ \sum_{s_j}\left(\frac{1}{2\delta} D\left(-a_j - \sqrt{\delta\varphi}\right) + \frac{\sqrt{\delta\varphi}}{\delta} F\left(-a_j - \sqrt{\delta\varphi}\right)\right)\end{array}\right)_{\sqrt{\delta\varphi} \geq b_j} +$$

$$\frac{1}{N}\sum_{j=1}^{N}\left(\begin{array}{l}\left(1 - \sum_{s_j} E\left(-a_j - \frac{b_j^2 + \delta\varphi}{2b_j}\right)\right)\ln\sum_{s_j}\nu + \\ \sum_{s_j}\left(\frac{1}{2\delta} D\left(-a_j - b_j\right) + \frac{b_j}{\delta} F\left(-a_j - \frac{b_j^2 + \delta\varphi}{2b_j}\right)\right)\end{array}\right)_{\sqrt{\delta\varphi} < b_j} \quad \text{(A26)}$$

$$a_j = \frac{2 f_j s_j z + \eta}{2\sqrt{f_j\left(1 - f_j\right)}}; \quad b_j = \sqrt{f_j\left(1 - f_j\right)}\tilde{\Delta}_j \delta\sqrt{t}$$

In this limit, the system of 9 saddle-point equations transformed into,

$$
\begin{aligned}
& f_{out} F\left(u_{-}\right)-\left(1-f_{out}\right) F\left(u_{+}\right)=0 \\
& f_{out} D\left(u_{-}\right)+\left(1-f_{out}\right) D\left(u_{+}\right)=\frac{4\tilde{\kappa}^2}{\alpha_c\left(u_{+}+u_{-}\right)^2} \varepsilon^2 t \\
& f_{out} F\left(u_{-}\right)+\left(1-f_{out}\right) F\left(u_{+}\right)=\frac{8\tilde{\kappa}^2}{\alpha_c\left(u_{+}+u_{-}\right)^3} \varepsilon t(\varepsilon-\delta) \\
& \frac{1}{N} \sum_{j=1}^N \sum_{s_j}\left(E\left(-a_j-\sqrt{\delta\varphi}\right)_{\sqrt{\delta\varphi} \geq b_j}+E\left(-a_j-\frac{b_j^2+\delta\varphi}{2 b_j}\right)_{\sqrt{\delta\varphi}<b_j}\right)=p \\
& \frac{1}{N} \sum_{j=1}^N \sum_{s_j} \frac{1}{\sqrt{f_j\left(1-f_j\right)}}\left(\begin{array}{l}
\left(F\left(-a_j-\sqrt{\delta\varphi}\right)+2 \sqrt{\delta\varphi} E\left(-a_j-\sqrt{\delta\varphi}\right)\right)_{\sqrt{\delta\varphi} \geq b_j}+ \\
\left(F\left(-a_j-b_j\right)+2 b_j E\left(-a_j-\frac{b_j^2+\delta\varphi}{2 b_j}\right)\right)_{\sqrt{\delta\varphi}<b_j}
\end{array}\right)=2 \delta \sqrt{t} \\
& \frac{1}{N} \sum_{j=1}^N \sum_{s_j}\left(E\left(-a_j-\sqrt{\delta\varphi}\right)_{\sqrt{\delta\varphi} \geq b_j}+E\left(-a_j-\frac{b_j^2+\delta\varphi}{2 b_j}\right)_{\sqrt{\delta\varphi}<b_j}\right)=1-\frac{\tilde{w}_x}{\nu} \\
& \frac{1}{N} \sum_{j=1}^N \sum_{s_j} \frac{f_j s_j}{\sqrt{f_j\left(1-f_j\right)}}\left(\begin{array}{l}
\left(F\left(-a_j-\sqrt{\delta\varphi}\right)+2 \sqrt{\delta\varphi} E\left(-a_j-\sqrt{\delta\varphi}\right)\right)_{\sqrt{\delta\varphi} \geq b_j}+ \\
\left(F\left(-a_j-b_j\right)+2 b_j E\left(-a_j-\frac{b_j^2+\delta\varphi}{2 b_j}\right)\right)_{\sqrt{\delta\varphi}<b_j}
\end{array}\right)=2 \tilde{h} \delta \sqrt{t} \\
& \frac{1}{N} \sum_{j=1}^N \sum_{s_j}\left(\begin{array}{l}
\left(D\left(-a_j-\sqrt{\delta\varphi}\right)+2 \sqrt{\delta\varphi} F\left(-a_j-\sqrt{\delta\varphi}\right)+2 \delta\varphi E\left(-a_j-\sqrt{\delta\varphi}\right)\right)_{\sqrt{\delta\varphi} \geq b_j} \\
+\left(D\left(-a_j-b_j\right)+2 b_j F\left(-a_j-b_j\right)+2 b_j^2 E\left(-a_j-\frac{b_j^2+\delta\varphi}{2 b_j}\right)\right)_{\sqrt{\delta\varphi}<b_j}
\end{array}\right)=\frac{4 \tilde{\kappa}^2 \delta^2 t}{\left(u_{+}+u_{-}\right)^2} \\
& \frac{1}{N} \sum_{j=1}^N \sum_{s_j}\left(b_j F\left(-a_j-b_j\right)-b_j F\left(-a_j-\frac{b_j^2+\delta\varphi}{2 b_j}\right)+\left(b_j^2-\delta\varphi\right) E\left(-a_j-\frac{b_j^2+\delta\varphi}{2 b_j}\right)\right)_{\sqrt{\delta\varphi}<b_j}= \\
& \left(2 \tilde{h} z+\eta+\frac{4 \tilde{\kappa}^2}{\left(u_{+}+u_{-}\right)^2}(\delta-\varepsilon) \sqrt{t}\right) \delta \sqrt{t}
\end{aligned}
\qquad \text{(A27)}
$$

Eliminating $\nu, \varepsilon$ and introducing new variables $x=\sqrt{\delta\varphi} \geq 0$ and $Q=\delta\sqrt{t}$, we arrived at the final set of 7 equations and 2 inequality constraints that determine the critical capacity.

$$a_j = \frac{2 f_j s_j z + \eta}{2\sqrt{f_j\left(1-f_j\right)}}; \quad b_j = \sqrt{f_j\left(1-f_j\right)}\tilde{\Delta}_j Q$$

$$\begin{cases} f_{out} F\left(u_-\right) - \left(1-f_{out}\right) F\left(u_+\right) = 0 \\ \frac{1}{N}\sum_{j=1}^{N}\sum_{s_j}\left( E\left(-a_j - x\right)_{x \ge b_j} + E\left(-a_j - \frac{b_j^2 + x^2}{2b_j}\right)_{x<b_j} \right) = p \\ \frac{1}{N}\sum_{j=1}^{N}\sum_{s_j}\frac{1}{\sqrt{f_j\left(1-f_j\right)}}\left( \begin{array}{l} \left(F\left(-a_j - x\right) + 2xE\left(-a_j - x\right)\right)_{x\ge b_j} + \\ \left(F\left(-a_j - b_j\right) + 2b_j E\left(-a_j - \frac{b_j^2 + x^2}{2b_j}\right)\right)_{x<b_j} \end{array} \right) = 2Q \\ \frac{1}{N}\sum_{j=1}^{N}\sum_{s_j}\frac{f_j s_j}{\sqrt{f_j\left(1-f_j\right)}}\left( \begin{array}{l} \left(F\left(-a_j - x\right) + 2xE\left(-a_j - x\right)\right)_{x\ge b_j} + \\ \left(F\left(-a_j - b_j\right) + 2b_j E\left(-a_j - \frac{b_j^2 + x^2}{2b_j}\right)\right)_{x<b_j} \end{array} \right) = 2\tilde{h}Q \\ \frac{1}{N}\sum_{j=1}^{N}\sum_{s_j}\left( \begin{array}{l} \left(D\left(-a_j - x\right) + 2xF\left(-a_j - x\right) + 2x^2 E\left(-a_j - x\right)\right)_{x\ge b_j} + \\ \left(D\left(-a_j - b_j\right) + 2b_j F\left(-a_j - b_j\right) + 2b_j^2 E\left(-a_j - \frac{b_j^2 + x^2}{2b_j}\right)\right)_{x<b_j} \end{array} \right) = \frac{4\tilde{\kappa}^2}{\left(u_+ + u_-\right)^2} Q^2 \\ \frac{1}{N}\sum_{j=1}^{N}\sum_{s_j}\left( b_j F\left(-a_j - b_j\right) - b_j F\left(-a_j - \frac{b_j^2 + x^2}{2b_j}\right) + \left(b_j^2 - x^2\right) E\left(-a_j - \frac{b_j^2 + x^2}{2b_j}\right)\right)_{x<b_j} = \\ = \left( 2\tilde{h}z + \eta - \frac{f_{out} F\left(u_-\right) + \left(1-f_{out}\right) F\left(u_+\right)}{f_{out} E\left(u_-\right) + \left(1-f_{out}\right) E\left(u_+\right)} \frac{2\tilde{\kappa}^2}{\left(u_+ + u_-\right)} Q \right) Q \\ x \ge 0; \quad u_+ + u_- \ge 0 \end{cases}$$

$$\alpha_c = \frac{f_{out} D\left(u_-\right) + \left(1-f_{out}\right) D\left(u_+\right)}{\left(f_{out} E\left(u_-\right) + \left(1-f_{out}\right) E\left(u_+\right)\right)^2} \frac{4\tilde{\kappa}^2}{\left(u_+ + u_-\right)^2} Q^2$$

(A28)

As expected, these equations did not contain the parameter $\tilde{w}_x$ introduced for convenience in Eqs (A6). When *sign* constraints are present, $s_j$ (which also appears in $a_j$) must be replaced with $g_j$ and the summation over $s_j$ must be removed.

**Distribution of input strengths at critical capacity**

The probability density function for connection *i* to have a specific sign, connectivity, and weight ($s,c,\tilde{w}$) was obtained from the following general expression.

$$p_i\left(s,c,\tilde{w}\right)=\left\langle \frac{1}{\Omega\left(\left\{X_j^\mu\right\},\left\{y^\mu\right\}\right)}\sum_{\left\{s_j,c_j\right\}}\delta_{s_i,s}\delta_{c_i,c}\int\prod_{j=1}^{N}d\tilde{w}_j\delta\left(\tilde{w}_i-\tilde{w}\right)\times \right.$$
$$\prod_{\mu=1}^{m}\theta\left(\left(2y^\mu-1\right)\left(\frac{1}{N}\sum_{j=1}^{N}s_jc_j\tilde{w}_jX_j^\mu-\tilde{h}\right)-\frac{\tilde{\kappa}}{\sqrt{N}}\right)\times \tag{A29}$$
$$\left.\prod_{j=1}^{N}\theta\left(\tilde{w}_j-\tilde{\Delta}_jc_j\right)\delta_{\sum_{j=1}^{N}c_j,Np}\,\delta\left(\sum_{j=1}^{N}c_j\tilde{w}_j-N\right)\delta\left(\sum_{j=1}^{N}\left(1-c_j\right)\tilde{w}_j-N\tilde{w}_x\right)\right\rangle_{X_j^\mu,y^\mu}$$

This expression was cast into a form amenable to analytic calculations by introducing *n* replica systems and taking the $n\to 0$ limit after averaging over the associations.

$$p_i\left(s,c,\tilde{w}\right)=\lim_{n\to 0}\left\langle \sum_{\left\{s_j^a,c_j^a\right\}}\delta_{s_i^1,s}\delta_{c_i^1,c}\int\prod_{a,j=1}^{n,N}d\tilde{w}_j^a\delta\left(\tilde{w}_i^1-\tilde{w}\right)\times \right.$$
$$\prod_{a,\mu=1}^{n,m}\theta\left(\left(2y^\mu-1\right)\left(\frac{1}{N}\sum_{j=1}^{N}s_j^ac_j^a\tilde{w}_j^aX_j^\mu-\tilde{h}\right)-\frac{\tilde{\kappa}}{\sqrt{N}}\right)\times \tag{A30}$$
$$\left.\prod_{a,j=1}^{n,N}\theta\left(\tilde{w}_j^a-\tilde{\Delta}_jc_j^a\right)\prod_{a=1}^{n}\delta_{\sum_{j=1}^{N}c_j^a,Np}\prod_{a=1}^{n}\delta\left(\sum_{j=1}^{N}c_j^a\tilde{w}_j^a-N\right)\prod_{a=1}^{n}\delta\left(\sum_{j=1}^{N}\left(1-c_j^a\right)\tilde{w}_j^a-N\tilde{w}_x\right)\right\rangle_{X_j^\mu,y^\mu}$$

Following the replica steps described above, we found,

$$p_i\left(s,c,\tilde{w}\right)=\int_{-\infty}^{\infty}\frac{e^{-x^2}}{\sqrt{\pi}}dx\frac{\delta_{c,0}e^{\frac{-\tilde{w}}{v}}+\delta_{c,1}\left(e^{\frac{1}{\delta}\left(2\frac{b_i}{\tilde{\Delta}_i}\left(x-a_i\right)\tilde{w}-\frac{b_i^2}{\tilde{\Delta}_i^2}\tilde{w}^2-\delta\varphi\right)}\right)_{\tilde{w}\geq\tilde{\Delta}_i}}{\sum_{s_i}\left(v+\sqrt{\frac{\pi}{4f_i\left(1-f_i\right)\delta t}}e^{\frac{1}{\delta}\left(x-a_i\right)^2}e^{-\varphi}\operatorname{erfc}\left(\frac{1}{\sqrt{\delta}}\left(-x+a_i+b_i\right)\right)\right)} \tag{A31}$$

This probability density function is normalized according to $\sum_{s,c}\int d'\tilde{w}p_i\left(s,c,\tilde{w}\right)=1$. At critical capacity, this expression was simplified by using Eqs. (A10, A11).

$$p_i\left(s,c,\tilde{w}\right)=\left[\delta_{c,0}\frac{\left(1-\sum_{s_i}E\left(-a_i-x\right)\right)}{\nu\sum_{s_i}\left|s_i\right|}e^{\frac{-\tilde{w}}{\nu}}+\delta_{c,1}\theta\left(\tilde{w}-\frac{\tilde{\Delta}_i x}{b_i}\right)\frac{b_i}{\sqrt{\pi}\tilde{\Delta}_i}e^{-\left(a_i+\frac{b_i}{\tilde{\Delta}_i}\tilde{w}\right)^2}\right]_{x\geq b_i}+$$

$$\left[\delta_{c,0}\frac{\left(1-\sum_{s_i}E\left(-a_i-\frac{b_i^2+x^2}{2b_i}\right)\right)}{\nu\sum_{s_i}\left|s_i\right|}e^{\frac{-\tilde{w}}{\nu}}+\right. \tag{A32}$$

$$\left.\delta_{c,1}\left(\left(E\left(a_i+b_i\right)-E\left(a_i+\frac{b_i^2+x^2}{2b_i}\right)\right)\delta\left(\tilde{w}-\tilde{\Delta}_i\right)+\frac{\theta\left(\tilde{w}-\tilde{\Delta}_i\right)b_i}{\sqrt{\pi}\tilde{\Delta}_i}e^{-\left(a_i+\frac{b_i}{\tilde{\Delta}_i}\tilde{w}\right)^2}\right)\right]_{x<b_i}$$

The probability density function for connection strength $\tilde{J}=\frac{J}{w}=sc\tilde{w}$ followed from this expression.

$$p_i\left(\tilde{J}\right)=\left[\left(1-\sum_{s_i}E\left(-a_i-x\right)\right)\delta\left(\tilde{J}\right)+\sum_{s_i}\theta\left(s_i\tilde{J}-\frac{\tilde{\Delta}_i x}{b_i}\right)\frac{b_i}{\sqrt{\pi}\tilde{\Delta}_i}e^{-\left(a_i+\frac{b_i}{\tilde{\Delta}_i}s_i\tilde{J}\right)^2}\right]_{x\geq b_i}+$$

$$\left[\left(1-\sum_{s_i}E\left(-a_i-\frac{b_i^2+x^2}{2b_i}\right)\right)\delta\left(\tilde{J}\right)+\right. \tag{A33}$$

$$\left.\sum_{s_i}\left(\left(E\left(a_i+b_i\right)-E\left(a_i+\frac{b_i^2+x^2}{2b_i}\right)\right)\delta\left(s_i\tilde{J}-\tilde{\Delta}_i\right)+\frac{\theta\left(s_i\tilde{J}-\tilde{\Delta}_i\right)b_i}{\sqrt{\pi}\tilde{\Delta}_i}e^{-\left(a_i+\frac{b_i}{\tilde{\Delta}_i}s_i\tilde{J}\right)^2}\right)\right]_{x<b_i}$$

Examples of probability density functions based on this equation are shown in Figure 3 of the main text. In general, these functions consist of two truncated Gaussians, one for $\tilde{J}>0$ and another for $\tilde{J}<0$, separated by identical gaps, $\left(\frac{x}{\sqrt{f_j\left(1-f_j\right)Q}}\right)_{x\geq b_i}+\left(\tilde{\Delta}_i\right)_{x<b_i}$, from a finite fraction of zero-weight connections represented by a Dirac delta function at $\tilde{J} = 0$. In the $x < b_i$ case, in which the gaps are equal to $\tilde{\Delta}_i$, there are additional finite fractions of connection strengths present at the outer boundaries of the gaps.

We defined sparsity, $S$, as the fraction of zero-weight connections of the constrained perceptron. Eq. (A33) yielded the following expression for this quantity.

$$S = 1 - \frac{1}{N}\sum_{i=1}^{N}\left(\left(\sum_{s_i} E\left(-a_i - x\right)\right)_{x \geq b_i} + \left(\sum_{s_i} E\left(-a_i - \frac{b_i^2 + x^2}{2b_i}\right)\right)_{x < b_i}\right) \qquad \text{(A34)}$$

**Replica solution of the model in the cases considered in the main text**

***h* + *κ* + *ℓ*1 case:** The solution in this case encompasses the $\ell_1$, $h + \ell_1$, and $\kappa + \ell_1$ cases of the main text. It was obtained from Eqs. (A28, A33, A34) by setting $\tilde{\Delta}_j = 0$ to eliminate the *gap* constraints and setting $x = 0$ to eliminate the $\ell_0$ norm constraint.

$$
\begin{cases}
f_{out} F\left(u_-\right) - \left(1 - f_{out}\right) F\left(u_+\right) = 0 \\
\dfrac{1}{N}\sum_{j=1}^{N} \dfrac{1}{\sqrt{f_j\left(1-f_j\right)}} \left( F\left( \dfrac{2 f_j z - \eta}{2\sqrt{f_j\left(1-f_j\right)}} \right) + F\left( \dfrac{-2 f_j z - \eta}{2\sqrt{f_j\left(1-f_j\right)}} \right) \right) = 2Q \\
\dfrac{1}{N}\sum_{j=1}^{N} \dfrac{f_j}{\sqrt{f_j\left(1-f_j\right)}} \left( -F\left( \dfrac{2 f_j z - \eta}{2\sqrt{f_j\left(1-f_j\right)}} \right) + F\left( \dfrac{-2 f_j z - \eta}{2\sqrt{f_j\left(1-f_j\right)}} \right) \right) = 2\tilde{h} Q \\
\dfrac{1}{N}\sum_{j=1}^{N} \left( D\left( \dfrac{2 f_j z - \eta}{2\sqrt{f_j\left(1-f_j\right)}} \right) + D\left( \dfrac{-2 f_j z - \eta}{2\sqrt{f_j\left(1-f_j\right)}} \right) \right) = \dfrac{4\tilde{\kappa}^2}{\left(u_+ + u_-\right)^2} Q^2 \\
Q = \left(2\tilde{h} z + \eta\right) \dfrac{\left(u_+ + u_-\right)}{2\tilde{\kappa}^2} \dfrac{f_{out} E\left(u_-\right) + \left(1 - f_{out}\right) E\left(u_+\right)}{f_{out} F\left(u_-\right) + \left(1 - f_{out}\right) F\left(u_+\right)} \\
u_+ + u_- \geq 0
\end{cases}
$$

$$
\alpha_c = \frac{f_{out} D\left(u_-\right) + \left(1 - f_{out}\right) D\left(u_+\right)}{\left(f_{out} E\left(u_-\right) + \left(1 - f_{out}\right) E\left(u_+\right)\right)^2} \frac{4\tilde{\kappa}^2}{\left(u_+ + u_-\right)^2} Q^2
$$

$$
p_i\left(\tilde{J}\right) = \left( 1 - E\left( \frac{2 f_i z - \eta}{2\sqrt{f_i\left(1-f_i\right)}} \right) - E\left( \frac{-2 f_i z - \eta}{2\sqrt{f_i\left(1-f_i\right)}} \right) \right) \delta\left(\tilde{J}\right) +
$$

$$
\frac{\sqrt{f_i\left(1-f_i\right)} Q}{\sqrt{\pi}} \left( \theta\left(-\tilde{J}\right) e^{-\left( \frac{2 f_j z - \eta}{2\sqrt{f_j\left(1-f_j\right)}} + \sqrt{f_i\left(1-f_i\right)} Q \tilde{J} \right)^2} + \theta\left(\tilde{J}\right) e^{-\left( \frac{2 f_j z + \eta}{2\sqrt{f_j\left(1-f_j\right)}} + \sqrt{f_i\left(1-f_i\right)} Q \tilde{J} \right)^2} \right)
$$

$$
S = 1 - \frac{1}{N}\sum_{i=1}^{N} \left( E\left( \frac{2 f_i z - \eta}{2\sqrt{f_i\left(1-f_i\right)}} \right) + E\left( \frac{-2 f_i z - \eta}{2\sqrt{f_i\left(1-f_i\right)}} \right) \right) \tag{A35}
$$

The solution for the **$\kappa$ + $\ell_1$ case** was obtained from Eqs. (A35) by setting $\tilde{h} = 0$.

The solution for the **$h$ + $\ell_1$ case** was obtained from Eqs. (A35) by taking the $\tilde{\kappa} \to 0$ limit, in which case $u_+ + u_- \to 0$, $2\tilde{h} z + \eta \to 0$, significantly simplifying the equations.

$$
\begin{cases}
f_{out}F\left(-u_{+}\right)-\left(1-f_{out}\right)F\left(u_{+}\right)=0 \\
\frac{1}{N}\sum_{j=1}^{N}\frac{f_j+\tilde{h}}{\sqrt{f_j\left(1-f_j\right)}}F\left(\frac{f_j+\tilde{h}}{\sqrt{f_j\left(1-f_j\right)}}z\right)+\frac{1}{N}\sum_{j=1}^{N}\frac{-f_j+\tilde{h}}{\sqrt{f_j\left(1-f_j\right)}}F\left(\frac{-f_j+\tilde{h}}{\sqrt{f_j\left(1-f_j\right)}}z\right)=0
\end{cases}
$$

$$
S=1-\frac{1}{N}\sum_{i=1}^{N}\left(E\left(\frac{f_j+\tilde{h}}{\sqrt{f_j\left(1-f_j\right)}}z\right)+E\left(\frac{-f_j+\tilde{h}}{\sqrt{f_j\left(1-f_j\right)}}z\right)\right) \tag{A36}
$$

$$
\alpha_c=\frac{1-S}{f_{out}E\left(-u_{+}\right)+\left(1-f_{out}\right)E\left(u_{+}\right)}
$$

The solution for the ***ℓ*1 case** was obtained from these expressions by setting $\tilde{h}=0$.

$$
\begin{aligned}
&f_{out}F\left(-u_{+}\right)-\left(1-f_{out}\right)F\left(u_{+}\right)=0 \\
&S=0 \\
&\alpha_c=\frac{1}{f_{out}E\left(-u_{+}\right)+\left(1-f_{out}\right)E\left(u_{+}\right)}
\end{aligned} \tag{A37}
$$

Lastly, when $f_{out}=0.5$, Eqs. (A37) yielded $\alpha_c=2$, which is a well-known result initially derived by Cover [6].

***h + κ + sign + ℓ*1 case:** In this case, we again set $\tilde{\Delta}_j=0$ to eliminate the *gap* constraints and *x* = 0 to eliminate the $\ell_0$ constraint in Eqs. (A28, A33, A34). In addition, we removed the sums over *s* and replaced *s* with *g*.

$$
\begin{cases}
f_{out}F\left(u_-\right)-\left(1-f_{out}\right)F\left(u_+\right)=0 \\
\frac{1}{N}\sum_{j=1}^{N}\frac{1}{\sqrt{f_j\left(1-f_j\right)}}F\left(\frac{-2f_jg_jz-\eta}{2\sqrt{f_j\left(1-f_j\right)}}\right)=2Q \\
\frac{1}{N}\sum_{j=1}^{N}\frac{f_jg_j}{\sqrt{f_j\left(1-f_j\right)}}F\left(\frac{-2f_jg_jz-\eta}{2\sqrt{f_j\left(1-f_j\right)}}\right)=2\tilde{h}Q \\
\frac{1}{N}\sum_{j=1}^{N}D\left(\frac{-2f_jg_jz-\eta}{2\sqrt{f_j\left(1-f_j\right)}}\right)=\frac{4\tilde{\kappa}^2}{\left(u_++u_-\right)^2}Q^2 \\
Q=\left(2\tilde{h}z+\eta\right)\frac{\left(u_++u_-\right)}{2\tilde{\kappa}^2}\frac{f_{out}E\left(u_-\right)+\left(1-f_{out}\right)E\left(u_+\right)}{f_{out}F\left(u_-\right)+\left(1-f_{out}\right)F\left(u_+\right)} \\
u_++u_-\geq 0
\end{cases}
$$

$$
\alpha_c=\frac{f_{out}D\left(u_-\right)+\left(1-f_{out}\right)D\left(u_+\right)}{\left(f_{out}E\left(u_-\right)+\left(1-f_{out}\right)E\left(u_+\right)\right)^2}\frac{4\tilde{\kappa}^2}{\left(u_++u_-\right)^2}Q^2
$$

$$
p_i\left(\tilde{J}\right)=\left(1-E\left(\frac{-2f_ig_iz-\eta}{2\sqrt{f_i\left(1-f_i\right)}}\right)\right)\delta\left(\tilde{J}\right)+\frac{\sqrt{f_i\left(1-f_i\right)}Q}{\sqrt{\pi}}\theta\left(g_i\tilde{J}\right)e^{-\left(\frac{2f_iz+g_i\eta}{2\sqrt{f_i\left(1-f_i\right)}}+\sqrt{f_i\left(1-f_i\right)}Q\tilde{J}\right)^2} \qquad \text{(A38)}
$$

$$
S=1-\frac{1}{N}\sum_{i=1}^{N}E\left(\frac{-2f_ig_iz-\eta}{2\sqrt{f_i\left(1-f_i\right)}}\right)
$$

This solution is consistent with our previously published result [16,29]. In the $\tilde{\kappa}\rightarrow 0$ limit, $u_++u_-\rightarrow 0$, $2\tilde{h}z+\eta\rightarrow 0$, and Eqs. (A38) could be simplified.

***h* + *κ* + *ℓ*0 + *ℓ*1 case:** We set $\tilde{\Delta}_j=0$ in Eqs. (A28, A33, A34) to eliminate the *gap* constraints.

$$
\begin{cases}
f_{out}F(u_-)-(1-f_{out})F(u_+)=0 \\
\frac{1}{N}\sum_{j=1}^{N}\sum_{s_j}E\left(\frac{-2f_js_jz-\eta}{2\sqrt{f_j(1-f_j)}}-x\right)=p \\
\frac{1}{N}\sum_{j=1}^{N}\sum_{s_j}\frac{1}{\sqrt{f_j(1-f_j)}}\left(F\left(\frac{-2f_js_jz-\eta}{2\sqrt{f_j(1-f_j)}}-x\right)+2xE\left(\frac{-2f_js_jz-\eta}{2\sqrt{f_j(1-f_j)}}-x\right)\right)=2Q \\
\frac{1}{N}\sum_{j=1}^{N}\sum_{s_j}\frac{f_js_j}{\sqrt{f_j(1-f_j)}}\left(F\left(\frac{-2f_js_jz-\eta}{2\sqrt{f_j(1-f_j)}}-x\right)+2xE\left(\frac{-2f_js_jz-\eta}{2\sqrt{f_j(1-f_j)}}-x\right)\right)=2\tilde{h}Q \\
\frac{1}{N}\sum_{j=1}^{N}\sum_{s_j}\left(\begin{array}{l} D\left(\frac{-2f_js_jz-\eta}{2\sqrt{f_j(1-f_j)}}-x\right)+2xF\left(\frac{-2f_js_jz-\eta}{2\sqrt{f_j(1-f_j)}}-x\right) \\ +2x^2E\left(\frac{-2f_js_jz-\eta}{2\sqrt{f_j(1-f_j)}}-x\right)\end{array}\right)=\frac{4\tilde{\kappa}^2}{(u_++u_-)^2}Q^2 \\
Q=\left(2\tilde{h}z+\eta\right)\frac{(u_++u_-)}{2\tilde{\kappa}^2}\frac{f_{out}E(u_-)+(1-f_{out})E(u_+)}{f_{out}F(u_-)+(1-f_{out})F(u_+)} \\
x\geq 0;\quad u_++u_-\geq 0
\end{cases}
$$

$$
\alpha_c=\frac{f_{out}D(u_-)+(1-f_{out})D(u_+)}{\left(f_{out}E(u_-)+(1-f_{out})E(u_+)\right)^2}\frac{4\tilde{\kappa}^2}{(u_++u_-)^2}Q^2
$$

$$
p_i(\tilde{J})=\left(1-\sum_{s_i}E\left(\frac{-2f_is_iz-\eta}{2\sqrt{f_i(1-f_i)}}-x\right)\right)\delta(\tilde{J})+ \tag{A39}
$$

$$
\sum_{s_i}\theta\left(s_i\tilde{J}-\frac{x}{\sqrt{f_i(1-f_i)}Q}\right)\frac{\sqrt{f_i(1-f_i)}Q}{\sqrt{\pi}}e^{-\left(\frac{2f_iz+s_i\eta}{2\sqrt{f_i(1-f_i)}}+\sqrt{f_i(1-f_i)}Q\tilde{J}\right)^2}
$$

$$
S=1-p
$$

Again, in the absence of robustness $u_++u_-\to 0$, $2\tilde{h}z+\eta\to 0$, and these equations could be simplified.

***h* + *κ* + *gap* + *ℓ*1 case:** This case corresponds to $x = 0$.

$$
\begin{cases}
f_{out}F(u_-)-(1-f_{out})F(u_+)=0 \\
\frac{1}{N}\sum_{j=1}^{N}\sum_{s_j}\frac{1}{\sqrt{f_j(1-f_j)}}\left(F(-a_j-b_j)+2b_jE\left(-a_j-\frac{b_j}{2}\right)\right)=2Q \\
\frac{1}{N}\sum_{j=1}^{N}\sum_{s_j}\frac{f_js_j}{\sqrt{f_j(1-f_j)}}\left(F(-a_j-b_j)+2b_jE\left(-a_j-\frac{b_j}{2}\right)\right)=2\tilde{h}Q \\
\frac{1}{N}\sum_{j=1}^{N}\sum_{s_j}\left(D(-a_j-b_j)+2b_jF(-a_j-b_j)+2b_j^2E\left(-a_j-\frac{b_j}{2}\right)\right)=\frac{4\tilde{\kappa}^2}{(u_++u_-)^2}Q^2 \\
\frac{1}{N}\sum_{j=1}^{N}\sum_{s_j}\left(b_jF(-a_j-b_j)-b_jF\left(-a_j-\frac{b_j}{2}\right)+b_j^2E\left(-a_j-\frac{b_j}{2}\right)\right)= \\
=\left(2\tilde{h}z+\eta-\frac{f_{out}F(u_-)+(1-f_{out})F(u_+)}{f_{out}E(u_-)+(1-f_{out})E(u_+)}\frac{2\tilde{\kappa}^2}{(u_++u_-)}Q\right)Q \\
u_++u_-\geq 0
\end{cases}
$$

$$
\alpha_c=\frac{f_{out}D(u_-)+(1-f_{out})D(u_+)}{\left(f_{out}E(u_-)+(1-f_{out})E(u_+)\right)^2}\frac{4\tilde{\kappa}^2}{(u_++u_-)^2}Q^2
$$

$$
p_i(\tilde{J})=\left(1-\sum_{s_i}E\left(-a_i-\frac{b_i}{2}\right)\right)\delta(\tilde{J})+ \tag{A40}
$$

$$
\sum_{s_i}\left(\left(E(a_i+b_i)-E\left(a_i+\frac{b_i}{2}\right)\right)\delta\left(s_i\tilde{J}-\tilde{\Delta}_i\right)+\frac{\theta\left(s_i\tilde{J}-\tilde{\Delta}_i\right)b_i}{\sqrt{\pi}\tilde{\Delta}_i}e^{-\left(a_i+\frac{b_i}{\tilde{\Delta}_i}s_i\tilde{J}\right)^2}\right)
$$

$$
S=1-\frac{1}{N}\sum_{i=1}^{N}\sum_{s_i}E\left(-a_i-\frac{b_i}{2}\right)
$$

Eqs. (A35-A40) were solved numerically in MATLAB to produce the results of the main text. The code is available at https://github.com/neurogeometry/Sparse_Learning.

**Validation of analytical results with numerical simulations**

To validate the results of replica calculations, Eqs. (A6) were also solved numerically. To that end, the model was cast into an optimization problem amenable to linear and mixed-integer linear programming and solved in MATLAB. The code is available at

https://github.com/neurogeometry/Sparse_Learning. Below, we briefly explain how the model was transformed into an optimization problem in some of the considered cases.

***h* + *κ* + *ℓ*1 case:** The model was solved by maximizing $\tilde{\kappa}$ for a given memory load *m*.

$$\begin{aligned}&\max(\tilde{\kappa})\\&(2y^{\mu}-1)\left(\frac{1}{N}\sum_{j=1}^{N}\tilde{J}_jX_j^{\mu}-\tilde{h}\right)>\frac{\tilde{\kappa}}{\sqrt{N}},\quad \mu=1,\ldots,m\\&\sum_{j=1}^{N}\left|\tilde{J}_j\right|-N=0\end{aligned} \tag{A41}$$

This problem was linearized with the following substitution and solved with the *linprog.m* function.

$$\tilde{J}_j=\tilde{J}_{j1}-\tilde{J}_{j2};\quad \tilde{J}_{j1},\tilde{J}_{j2}\geq 0;\quad \left|\tilde{J}_j\right|=\tilde{J}_{j1}+\tilde{J}_{j2} \tag{A42}$$

***h* + *κ* + *sign* + *ℓ*1 case:** This problem was also solved by maximizing $\tilde{\kappa}$. It is already linear because $\left|\tilde{J}_j\right|=g_j\tilde{J}_j$.

$$\begin{aligned}&\max(\tilde{\kappa})\\&(2y^{\mu}-1)\left(\frac{1}{N}\sum_{j=1}^{N}\tilde{J}_jX_j^{\mu}-\tilde{h}\right)>\frac{\tilde{\kappa}}{\sqrt{N}},\quad \mu=1,\ldots,m\\&g_j\tilde{J}_j\geq 0,\quad j=1,\ldots,N\\&\sum_{j=1}^{N}g_j\tilde{J}_j-N=0\end{aligned} \tag{A43}$$

***h* + *κ* + *ℓ*0 + *ℓ*1 case:** The model in this case was transformed into a mixed-integer linear programming problem by using the following substitution.

$$\tilde{J}_j=c_jy_j;\quad c_j=\{0,1\};\quad y_j=y_{1j}-y_{2j};\quad y_{1j},y_{2j}\geq 0;\quad \left|y_j\right|=y_{1j}+y_{2j} \tag{A44}$$

This led to a problem that is suited for the *intlinprog.m* function.

$$
\begin{aligned}
&\max(\tilde{\kappa}) \\
&-Nc_j \le y_{1j} - y_{2j} \le Nc_j, \quad j = 1,\ldots,N \\
&y_{1j}, y_{2j} \ge 0, \quad j = 1,\ldots,N \\
&\left(2y^{\mu} - 1\right)\left(\frac{1}{N}\sum_{j=1}^{N}\left(y_{1j} - y_{2j}\right)X_j^{\mu} - \tilde{h}\right) > \frac{\tilde{\kappa}}{\sqrt{N}}, \quad \mu = 1,\ldots,m \\
&\sum_{j=1}^{N} c_j - Np = 0 \\
&\sum_{j=1}^{N}\left(y_{1j} + y_{2j}\right) - N = 0
\end{aligned}
\tag{A45}
$$

***h* + *κ* + *gap* + *ℓ*1 case:** The problem was linearized and solved numerically with *intlinprog.m* by utilizing the binary variables, $\left\{c_j, s_j\right\}$.

$$
\begin{aligned}
&\max(\tilde{\kappa}) \\
&\tilde{w}_j + N\left(s_j - 1\right) \le \tilde{J}_j \le -\tilde{w}_j + N\left(s_j + 1\right), \quad j = 1,\ldots,N \\
&-\tilde{w}_j \le \tilde{J}_j \le \tilde{w}_j, \quad j = 1,\ldots,N \\
&\tilde{\Delta}_j c_j \le \tilde{w}_j \le Nc_j, \quad j = 1,\ldots,N \\
&\left(2y^{\mu} - 1\right)\left(\frac{1}{N}\sum_{j=1}^{N}\tilde{J}_j X_j^{\mu} - \tilde{h}\right) > \frac{\tilde{\kappa}}{\sqrt{N}}, \quad \mu = 1,\ldots,m \\
&\sum_{j=1}^{N}\tilde{w}_j - N = 0
\end{aligned}
\tag{A46}
$$

Results of numerical simulations based on Eqs. (A41-A46) are shown in Figure S1 together with the corresponding analytical results of replica calculations.

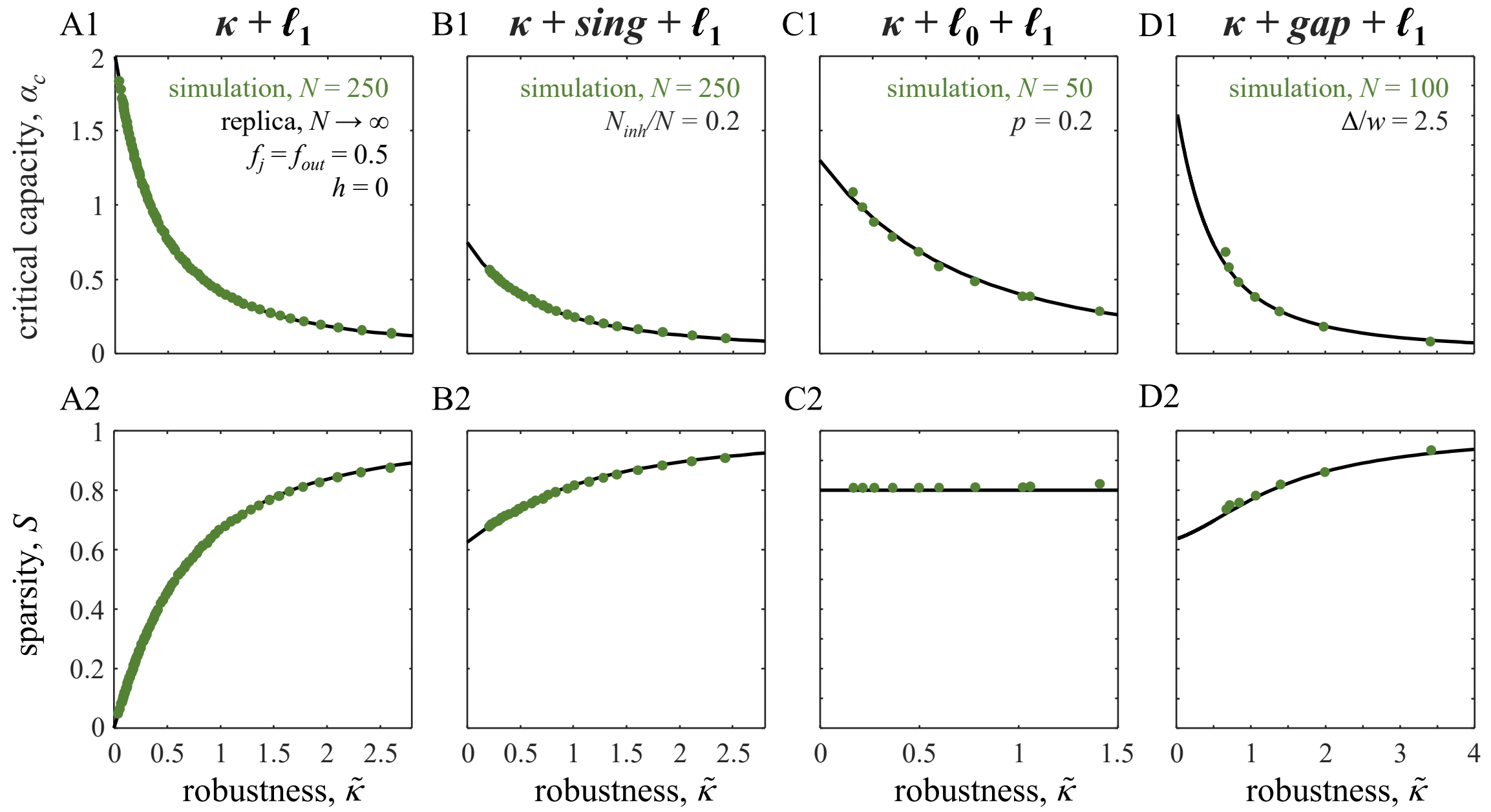

**Figure S1:** Validation of theoretical results with numerical simulations. Critical capacity (**A1**) and sparsity (**A2**) as functions of robustness in the $\kappa + \ell_1$ constrained case. Model parameters are displayed in (A1). Results of numerical simulations for $N = 250$ inputs (green points) are in good agreement with the results based on replica theory (black lines). Numerical results were obtained with methods of linear optimization and averaged over 100 runs for every parameter setting. **B1,2**. Same for the $\kappa + sign + \ell_1$ case with the fraction of inhibitory inputs $N_{inh}/N = 0.2$. **C1,2**. Same in the $\kappa + \ell_0 + \ell_1$ case with $p = 0.2$. In this case, $N = 50$ was used due to the NP-hard nature of the problem. **D1,2**. Same in the $\kappa + gap + \ell_1$ case with $N = 100$ and the gap set to $\Delta/w = 2.5$.

## Sparse learning can be done online with a perceptron-type learning rule

To solve the constrained associative learning problem of Eqs. (A2) online, which may be necessary for computational neuroscience and ML applications, we introduced a sparse learning rule based on perceptron learning [18]. Here, the connection strengths are modified stochastically in response to associations that are presented one at a time (or in batches) while the model attempts to learn a set of associations that can in general be unfeasible. The rule implements sparse learning in the presence of a fixed threshold $h$, $\kappa$, $\ell_1$, *sign*, and *gap* constraints. The $\ell_0$ constraint was excluded from this list because there is no simple way to implement it online due to its NP-hard nature [20].

The sparse learning rule works as follows. At every step, a single not yet robustly learned association is chosen at random, and the connection strengths of the perceptron are updated synchronously in five consecutive steps:

$$\text{choose } \mu: \left(2y^{\mu}-1\right)\left(\sum_{j=1}^{N} J_j X_j^{\mu} - h\right) \le \kappa$$

$$\text{update } J:$$

$$\begin{cases} J_j \mapsto J_j + \beta_{\mu}\left(2y^{\mu}-1\right)X_j^{\mu} \\ J_j \mapsto J_j \theta\left(J_j g_j\right) \\ J_j \mapsto J_j + \beta_w \operatorname{sgn}\left(J_j\right)\left(w - \frac{1}{N}\sum_{j=1}^{N}\left|J_j\right|\right) \\ J_j \mapsto J_j \theta\left(J_j g_j\right) \\ J_j \mapsto \Delta_j \operatorname{sgn}\left(J_j\right)\theta\left(r_j - p_{\Delta}\right),\ r_j \sim U\left(0,1\right),\ \text{if } 0 < \left|J_j\right| < \Delta_j \end{cases} \tag{A47}$$

Unlike the standard perceptron learning rule [18], the individual steps of Eqs. (A47) attempt to enforce the $\kappa$, $\ell_1$, *sign*, and *gap* constraints during learning. The first update step in Eqs. (A47) is a standard perceptron learning step, in which parameter $\beta_\mu$ controls the rate at which associations are learned. This step implements the $\kappa$ constraint by updating connection strengths for not robustly learned association [first line in Eqs. (A47)]. The second step enforces the *sign* constraints. The next two steps combined reduce the deviation of the $\ell_1$ norm from its target $Nw$ while enforcing the *sign* constraints, in which the former is controlled by the parameter $\beta_w$. Subsequently, connection strengths that lie in their gap regions are stochastically set to the gaps' boundaries, i.e. 0 with probability $p_\Delta$ and $\Delta_j \operatorname{sgn}\left(J_j\right)$ with probability $1 - p_\Delta$. The above sparse learning rule is not guaranteed to find solutions to feasible learning problems. It was designed to produce "good enough" or approximate solutions to learning problems that are not necessarily feasible, such as the ones encountered in ML applications and arguably those faced by the brain.

The sparse learning rule was evaluated on the MNIST dataset of handwritten digits [26]. Ten constrained perceptron models were trained to classify the ten digits. Each model was trained on 60,000 and tested on 10,000 examples. For the model of a given digit, all images of that digit were associated with $y = 1$ and the rest with $y = 0$. The training was run for $10^6$ sparse learning steps on randomly chosen examples with the learning parameters set at $\beta_\mu = 0.01$, $\beta_w = 0.1$, and $p_\Delta = 0.95$. Because the binary classification problem solved by every constrained perceptron is highly unbalanced (Positive/Negative examples ~ 1/9 ratio), the Balanced Accuracy [27] was used to evaluate the classification performance on training and testing examples. The results were averaged over all ten digits. In the cases involving *sign* and $q\ell_0$, the results were also averaged over 10 random initializations of these constraints.

MATLAB implementation of the sparse learning rule is available at https://github.com/neurogeometry/Sparse_Learning.